\documentclass[a4paper,twoside,reqno,11pt,tbtags]{article}

\usepackage{amssymb,amsmath,amsthm,mathrsfs}
\usepackage{graphicx}
\usepackage{times}

\leftmargini=\baselineskip

\newtheoremstyle{localthm}
	{5pt} 
	{5pt} 
	{\sl} 
	{} 
	{\bf} 
	{{\rm.}} 
	{.7em} 
	{} 

\theoremstyle{localthm}
\newtheorem{Theorem}{Theorem}
\newtheorem{Corollary}[Theorem]{Corollary}

\newtheorem{Lemma}[Theorem]{Lemma}

\newtheoremstyle{localrem}
	{5pt} 
	{5pt} 
	{\rm} 
	{} 
	{\bf} 
	{{\rm.}} 
	{.7em} 
	{} 

\theoremstyle{localrem}
\newtheorem{Remark}[Theorem]{Remark}

\newcommand{\Ex}{\mathop{\mathrm{I\!E}}\nolimits}

\newcommand{\R}{\mathbb{R}}

\newcommand{\LL}{\mathcal{L}}
\newcommand{\NN}{\mathcal{N}}

\newcommand{\argmin}{\mathop{\mathrm{arg\,min}}}

\newcommand{\diag}{\mathop{\mathrm{diag}}\nolimits}
\newcommand{\tr}{\mathop{\mathrm{tr}}\nolimits}

\newcommand{\bs}{\boldsymbol}

\newcommand{\bmu}{\bs{\mu}}
\newcommand{\bSigma}{\bs{\Sigma}}

\newcommand{\what}{\widehat}

\newcommand{\Rqq}{\R_{}^{q\times q}}

\newcommand{\Rqqsym}{\R_{\rm sym}^{q\times q}}
\newcommand{\Rqqsympd}{\R_{{\rm sym},>0}^{q\times q}}

\begin{document}

\title{New Algorithms for $M$-Estimation of\\
	Multivariate Scatter and Location}
\author{Lutz D\"umbgen$^{1*}$, Klaus Nordhausen$^{2**}$ and Heike Schuhmacher$^1$\\
	($^1$University of Bern and  $^2$University of Turku)}
\date{May 2015, revised October 2015}
\maketitle

\begin{abstract}
We present new algorithms for $M$-estimators of multivariate scatter and location and for symmetrized $M$-estimators of multivariate scatter. The new algorithms are considerably faster than currently used fixed-point and other algorithms. The main idea is to utilize a Taylor expansion of second order of the target functional and devise a partial Newton-Raphson procedure. In connection with symmetrized $M$-estimators we work with incomplete $U$-statistics to accelerate our procedures initially.
\end{abstract}

\vfill

\noindent
$^*$Work supported by Swiss National Science Foundation.\\
$^{**}$Work supported by Academy of Finland (grant 268703).



\paragraph{Corresponding author:}
Lutz D\"umbgen, e-mail: {\tt duembgen@stat.unibe.ch}

\addtolength{\baselineskip}{0.5\baselineskip}

\section{Introduction}
\label{sec:introduction}

Robust estimation of multivariate location and scatter for a distribution $P$ on $\R^q$ is a recurring topic in statistics. For instance, different estimators of multivariate scatter are an important ingredient for independent component analysis (ICA) or invariant coordinate selection (ICS), see Nordhausen et al.~\cite{Nordhausen_etal_2008} and Tyler et al.~\cite{Tyler_etal_2009} and the references therein. Of particular interest are $M$-estimators and their symmetrized versions as defined in Sections~\ref{subsec:ScatterOnly} and \ref{subsec:SymmetrizedM}, respectively, because they offer a good compromise between robustness and computational feasibility. The most popular algorithm to compute $M$-estimators of multivariate scatter is to iterate a fixed-point equation, see Huber~\cite{Huber_1981} (Section 8.11), Tyler~\cite{Tyler_1987a} and Kent and Tyler~\cite{Kent_Tyler_1991}. This algorithm has nice properties such as guaranteed convergence for any starting point. However, as discussed later, it can be rather slow for high dimensions and large data sets. We introduce two alternative methods, a gradient descent method with approximately optimal stepsize and a partial Newton-Raphson method, which turn out to be substantially faster.

Computation time becomes a major issue in connection with symmetrized $M$-estimators. These estimators are important because of a desirable ``block independence property'' as explained in Section~\ref{subsec:SymmetrizedM}; see also D\"umbgen~\cite{Duembgen_1998} and Sirki\"a et al.~\cite{Sirkiae_etal_2007}. If applied to a sample of $n$ observations $X_1, X_2, \ldots, X_n \in \R^q$, symmetrized $M$-estimators utilize the empirical distribution of all $\binom{n}{2}$ differences $X_i - X_j$, $1 \le i < j \le n$.

In Section~\ref{sec:M-functional} we describe briefly the various $M$-estimators we are interested in. Then we introduce a general target functional on the space of symmetric and positive definite matrices in $\R^{q\times q}$ which has to be minimized. Section~\ref{sec:analysis} presents some analytical properties of the latter functional which are essential to understand existing algorithms and to devise new ones. These parts follow closely a recent survey of multivariate $M$-functionals by D\"umbgen et al.~\cite{Duembgen_etal_2015}. In Section~\ref{sec:algorithms} we discuss the aforementioned fixed-point algorithm of Kent and Tyler~\cite{Kent_Tyler_1991} and explain rigorously why it is suboptimal. Then we introduce two alternative methods, a gradient descent method with approximately optimal stepsize and a partial Newton-Raphson method. Numerical experiments in Section~\ref{sec:Examples} show that the new algorithms are substantially faster than the fixed-point algorithms or the algorithms by Arslan et al.~\cite{Arslan_etal_1995}. Proofs are deferred to Section~\ref{sec:proofs}.

\paragraph{Some Notation.}
The space of symmetric matrices in $\R^{q\times q}$ is denoted by $\Rqqsym$, and $\Rqqsympd$ stands for its subset of positive definite matrices. The identity matrix in $\Rqq$ is written as $I_q$. The Euclidean norm of a vector $v \in \R^q$ is denoted by $\|v\| = \sqrt{v^\top v}$. For matrices $M, N$ with identical dimensions we write
\[
	\langle M, N\rangle \ := \ \tr(M^\top N)
	\quad\text{and}\quad
	\|M\| \ := \ \sqrt{\langle M, M\rangle} ,
\]
so $\|M\|$ is the Frobenius norm of $M$.

\section{The $M$-estimators and the target functional}
\label{sec:M-functional}

Let $X_1, X_2, \ldots, X_n$ be independent random vectors with unknown distribution $P$ on $\R^q$. Our task is to define and then estimate a certain center $\bs{\mu}(P) \in \R^q$ and scatter matrix $\bs{\Sigma}(P) \in \Rqqsympd$.

\subsection{The scatter-only problem}
\label{subsec:ScatterOnly}

Let us start with the assumption that $\bs{\mu}(P) = 0$. To define and estimate a scatter functional $\bs{\Sigma}(P)$ we consider a simple working model consisting of elliptically symmetric probability densities $f_\Sigma$ on $\R^q$ depending on a parameter $\Sigma \in \Rqqsympd$:
\[
	f_{\Sigma}(x)
	\ = \ C^{-1} \det(\Sigma)^{-1/2} \exp \bigl( - \rho(x^\top \Sigma^{-1}x)/2 \bigr) ,
\]
where $\rho : [0,\infty) \to \R$ is a given function such that $C :=  \int \exp \bigl( - \rho(\|x\|^2)/2 \bigr) \, dx$ is finite. Assuming temporarily that this working model is correct, one could estimate the true underlying matrix parameter by a maximizer of the corresponding log-likelihood function for this model,
\[
	\Sigma \ \mapsto \ - n \log C
		- \frac{1}{2} \sum_{i=1}^n \rho(X_i^\top \Sigma^{-1} X_i^{})
		- \frac{n}{2} \log \det(\Sigma) .
\]
With the empirical distribution $\what{P} = n^{-1} \sum_{i=1}^n \delta_{X_i}$ of the data $X_1, X_2, \ldots, X_n$, the log-likelihood at $\Sigma$ may be written as $n \int \log f_{\Sigma} \, d\what{P}$. Thus maximization of the log-likelihood function over $\Rqqsympd$ is equivalent to minimization of $\Sigma \mapsto L(\Sigma,\what{P})$, where
\begin{align*}
	L(\Sigma,Q) \
	&:= \ 2 \int \log(f_{I_q}/f_{\Sigma}) \, dQ \\
	&= \ \int \bigl[ \rho(x^\top \Sigma^{-1} x) - \rho(x^\top x) \bigr] \, Q(dx)
		+ \log \det(\Sigma)
\end{align*}
for a generic distribution $Q$ on $\R^q$. We include $f_{I_q}$ and $\rho(x^\top x)$, respectively, because often this increases the range of distributions $Q$ such that $L(\Sigma,Q)$ is well-defined in $\R$. If $L(\cdot,Q)$ has a unique maximizer over $\Rqqsympd$, we denote it with $\bs{\Sigma}(Q)$. The resulting mapping $Q \mapsto \bSigma(Q)$ is called an $M$-functional of scatter. In particular, $\bs{\Sigma}(\what{P})$ serves as an estimator of the scatter parameter $\bs{\Sigma}(P)$, assuming that both exist. If $P$ happens to have a density $f_{\Sigma_o}$ in our working model, then $\bs{\Sigma}(P) = \Sigma_o$. If $P$ is merely elliptically symmetric with center $0$ and scatter matrix $\Sigma_o$, for instance, if it has a density $f$ of the form
\[
	f(x) \ = \ \det(\Sigma_o)^{-1/2} g_o(x^\top \Sigma_o^{-1} x)
\]
with $g_o : [0,\infty) \to [0,\infty)$, then at least $\bs{\Sigma}(P) = \gamma \Sigma_o$ for some $\gamma > 0$.

An important example are multivariate $t$ distributions with $\nu > 0$ degress of freedom. Here $\rho = \rho_{\nu,q}$ with
\begin{equation}
\label{eq:rho_nu}
	\rho_{\nu,q}(s) \ = \ (\nu + q) \log(\nu + s)
	\quad\text{for} \ s \ge 0 .
\end{equation}
Note that $\rho(x^\top \Sigma^{-1}x) - \rho(x^\top x)$ equals $(q + \nu) \log \bigl( (\nu + x^\top \Sigma^{-1} x)/(\nu + x^\top x) \bigr)$, a bounded and smooth function of $x \in \R^q$.

\subsection{The location-scatter problem}
\label{subsec:LocationScatter}

Now our working model consists of probability densities $f_{\mu,\Sigma}$ on $\R^q$ with parameters $\mu \in \R^q$ and $\Sigma \in \Rqqsympd$, namely,
\[
	f_{\mu,\Sigma}(x)
	\ = \ C^{-1} \det(\Sigma)^{-1/2}
		\exp \Bigl( - \rho \bigl( (x - \mu)^\top \Sigma^{-1} (x - \mu) \bigr)/2 \Bigr) .
\]
Here $(\bs{\mu}(P'),\bs{\Sigma}(P'))$ is defined as the minimizer of $2 \int \log(f_{0,I_q}/f_{\mu,\Sigma}) \, dP'$, where $P'$ stands for $P$ or $\what{P}$. But now we utilize a trick of Kent and Tyler~\cite{Kent_Tyler_1991} to get back to a scatter-only problem: With
\begin{equation}
\label{eq:augmentation}
	y = y(x)
	\ := \ \begin{bmatrix} x \\ 1 \end{bmatrix}
	\quad\text{and}\quad
	\Gamma = \Gamma(\mu,\Sigma)
	\ := \ \begin{bmatrix}
		\Sigma + \mu\mu^\top & \mu \\ \mu^\top & 1
	\end{bmatrix}
\end{equation}
we may write $\log \det(\Sigma) = \log \det(\Gamma)$ and
\[
	- 2 \log f_{\mu,\Sigma}(x)
	\ = \ -2 \log(C) + \rho(y^\top \Gamma^{-1} y - 1)
		+ \log \det(\Gamma) .
\]
Hence $2 \int \log(f_{0,I_q}/f_{\mu,\Sigma}) \, dP'$ equals
\[
	L(\Gamma,Q)
	\ = \ \int \bigl[ \rho(y^\top \Gamma^{-1} y - 1) - \rho(y^\top y - 1) \bigr]
		\, Q(dy)
		+ \log\det(\Gamma)
\]
with $Q := \LL(y(X'))$, where $X' \sim P'$. Consequently, if $\bs{\Gamma} \in \R^{(q+1)\times(q+1)}_{{\rm sym},>0}$ minimizes $L(\cdot,Q)$ under the constraint
\[
	\bs{\Gamma}_{q+1,q+1} \ = \ 1 ,
\]
then we may write
\[
	\bs{\Gamma} \ = \ \begin{bmatrix}
		\bs{\Sigma}(P') + \bs{\mu}(P')\bs{\mu}(P')^\top & \bs{\mu}(P') \\
		\bs{\mu}(P')^\top & 1
	\end{bmatrix} ,
\]
and $(\bs{\mu}(P'),\bs{\Sigma}(P'))$ solves the original minimization problem. The mappings $P' \mapsto \bs{\mu}(P')$ and $P' \mapsto \bs{\Sigma}(P')$ are called $M$-functional of location and $M$-functional of scatter, respectively.

In the special case of $\rho = \rho_{\nu,q}$ with $\nu \ge 1$ we have the identity
\[
	\rho_{\nu,q}(s-1) \ = \ \rho_{\nu-1,q+1}(s)
	\quad\text{for} \ s > 0,
\]
where we define
\begin{equation}
\label{eq:rho_0}
	\rho_{0,q}(s) \ := \ q \log(s)
	\quad\text{for} \ s > 0 .
\end{equation}
In case of $\nu > 1$ one can show that any minimizer $\bs{\Gamma}$ of $L(\cdot,Q)$ does satisfy the equation $\bs{\Gamma}_{q+1,q+1} = 1$, see \cite{Kent_Tyler_1991} and \cite{Kent_etal_1994}. In case of $\nu = 1$, which corresponds to multivariate Cauchy distributions, any minimizer $\bs{\Gamma}$ of $L(\cdot,Q)$ may be rescaled such that $\bs{\Gamma}_{q+1,q+1} = 1$. Thus in connection with multivariate $t$ distributions with $\nu \ge 1$ degrees of freedom, the location-scatter problem can be reduced to a scatter-only problem.

If $P$ has a density $f_{\mu_o,\Sigma_o}$ in our working model, then $(\bs{\mu}(P),\bs{\Sigma}(P)) = (\mu_o,\Sigma_o)$. If $P$ is just elliptically symmetric with center $\mu_o$ and scatter matrix $\Sigma_o$, for instance, if it has a density $f$ of the form
\[
	f(x) \ = \ \det(\Sigma_o)^{-1/2}
		g_o \bigl( (x - \mu_o)^\top \Sigma_o^{-1} (x - \mu_o) \bigr)
\]
with $g_o : [0,\infty) \to [0,\infty)$, then $\bs{\mu}(P) = \mu_o$ and $\bs{\Sigma}(P) = \gamma \Sigma_o$ for some $\gamma > 0$.

\subsection{Symmetrized $M$-functionals}
\label{subsec:SymmetrizedM}

Suppose that $P$ is (approximately) elliptically symmetric with unknown center $\mu_o$ and unknown scatter matrix $\Sigma_o$. In many situations one is only interested in the ``shape matrix'' $\det(\Sigma_o)^{-1/q} \Sigma_o$, i.e.\ a positive multiple of $\Sigma_o$ with determinant $1$. Examples are principal components, regression and correlation measures, where multiplying $\Sigma_o$ with a positive scalar has no impact. Then we may get rid of the nuisance location parameter $\mu_o$ by replacing $P$ with its symmetrization
\[
	P \!\ominus\! P \ := \ \LL(X' - X'')
	\quad\text{with independent} \ X',X'' \sim P .
\]
Indeed, $P \ominus P$ is (approximately) elliptically symmetric with center $0$ and the same shape matrix $\det(\Sigma_o)^{-1/q} \Sigma_o$. We may estimate $P \ominus P$ by the measure-valued $U$-statistic
\[
	\what{P \!\ominus\! P} \ := \ \binom{n}{2}^{-1}
		\sum_{1 \le i < j \le n} \delta_{X_i - X_j}^{} .
\]
Then, if we define $\bs{\Sigma}(Q)$ to be the minimizer of
\[
	\int \bigl[ \rho(x^\top \Sigma^{-1} x) - \rho(x^\top x) \bigr] \, Q(dx)
		+ \log \det(\Sigma)
\]
with respect to $\Sigma$, then the shape matrix of $\bs{\Sigma}(\what{P \!\ominus\! P})$ is a plausible estimator of the true shape matrix $\det(\Sigma_o)^{-1/q} \Sigma_o$. The mapping $P \mapsto \bs{\Sigma}(P \!\ominus\! P)$ is called a symmetrized $M$-functional of scatter.

This symmetrization has a second, even more important advantage: Consider an arbitrary distribution $P$, i.e.\ it may fail to be (approximately) elliptically symmetric. But suppose that a random vector $X \sim P$ may be written as $X = [X_1^\top, X_2^\top]^\top$ with independent subvectors $X_1 \in \R^{q(1)}, X_2 \in \R^{q(2)}$. Then $\bs{\Sigma}(P)$ is block-diagonal in the sense that
\[
	\bs{\Sigma}(P) \ = \ \begin{bmatrix}
		\bs{\Sigma}_{1}(P) & \bs{0} \\
		\bs{0} & \bs{\Sigma}_{2}(P)
	\end{bmatrix}
\]
with symmetric matrices $\bs{\Sigma}_{i}(P) \in \R_{\rm sym}^{q(i)\times q(i)}$. For a further discussion on the use of symmetrized scatter matrices in multivariate statistics see also Nordhausen and Tyler~\cite{Nordhausen_Tyler_2015}.

\subsection{The general settings}

Let $Q$ be a probability distribution on $\R^q$. Now we seek to minimize a certain target functional $L(\cdot,Q)$ on the space $\Rqqsympd$ of symmetric and positive definite matrices in $\Rqq$, where $L(\cdot,\cdot)$ and $Q$ have to satisfy certain conditions:

\noindent
\textbf{Setting 0.} \
We assume that $Q(\{0\}) = 0$, and for $\Sigma \in \Rqqsympd$ we define
\[
	L_0(\Sigma,Q)
	\ := \ q \int \log \Bigl( \frac{x^\top \Sigma^{-1}x}{x^\top x} \Bigr) \, Q(dx)
		+ \log \det(\Sigma) .
\]
Moreover, we assume that
\[
	Q(\mathbb{V}) \ < \ \frac{\dim(\mathbb{V})}{q}
\]
for any linear subspace $\mathbb{V}$ of $\R^q$ with $1 \le \dim(\mathbb{V}) < q$.

\noindent
\textbf{Setting 1.} \ Let $\rho : [0,\infty) \to \R$ be twice continuously differentiable such that $\rho' > 0 \ge \rho''$. Further we assume that $\psi(s) := s \rho'(s)$ satisfies the following two properties: $\psi' > 0$ and $q < \psi(\infty) := \lim_{s \to \infty} \psi(s) < \infty$. For $\Sigma \in \Rqqsympd$ we define
\[
	L_\rho(\Sigma,Q)
	\ := \ \int \bigl[ \rho(x^\top \Sigma^{-1}x) - \rho(x^\top x) \bigr] \, Q(dx)
		+ \log \det(\Sigma) .
\]
Moreover, we assume that
\[
	Q(\mathbb{V}) \ < \ \frac{\psi(\infty) - q + \dim(\mathbb{V})}{\psi(\infty)}
\]
for any linear subspace $\mathbb{V}$ of $\R^q$ with $0 \le \dim(\mathbb{V}) < q$.

Note that for $\nu > 0$, $\rho = \rho_{\nu,q}$ satisfies the conditions of Setting~1 with $\psi(s) = (\nu + q) s/(\nu + s)$. Hence $\psi(\infty) = \nu + q$, and $Q$ has to satisfy
\[
	Q(\mathbb{V}) \ < \ \frac{\nu + \dim(\mathbb{V})}{\nu + q}
\]
for proper linear subspaces $\mathbb{V}$ of $\R^q$.

Note also that Setting~0 is similar to Setting~1 if we define $\rho := \rho_{0,q}$ as in \eqref{eq:rho_0}. The main difference to Setting~1 is that $L_0(t\Sigma,Q) = L_0(\Sigma,Q)$ for arbitrary $\Sigma \in \Rqqsympd$ and $t > 0$. In what follows we often write $L(\Sigma,Q)$ for $L_0(\Sigma,Q)$ or $L_\rho(\Sigma,Q)$.

The assumptions on $\rho$ and $Q$ imply that the functional $L(\cdot,Q)$ has essentially a unique minimizer (see \cite{Kent_Tyler_1991}, \cite{Dudley_etal_2009} or \cite{Duembgen_etal_2015}):

\begin{Theorem}
\label{thm:existence}
In Setting~0 there exists a unique matrix $\bs{\Sigma}_0(Q) \in \Rqqsympd$ such that
\[
	L_0(\bs{\Sigma}_0(Q),Q) \le L_0(\cdot,Q)
	\quad\text{and}\quad
	\det(\bs{\Sigma}_0(Q)) \ = \ 1 .
\]

\noindent
In Setting~1 there exists a unique matrix $\bs{\Sigma}_\rho(Q) \in \Rqqsympd$ such that
\[
	L_\rho(\bs{\Sigma}_\rho(Q),Q) \ \le \ L_\rho(\cdot,Q) .
\]
\end{Theorem}

Coming back to the specific situation with independent random variables $X_1, X_2, \ldots, X_n$ with distribution $P$ on $\R^q$, the scatter estimators in Sections~\ref{subsec:ScatterOnly}, \ref{subsec:LocationScatter} and \ref{subsec:SymmetrizedM} correspond to the following choices of $Q$:
\begin{itemize}
\item \ $Q = \what{P} = n^{-1} \sum_{i=1}^n \delta_{X_i}$ \
	(Section~\ref{subsec:ScatterOnly});
\item \ $Q = n^{-1} \sum_{i=1}^n \delta_{y(X_i)}$ with dimension $q+1$ in place of $q$ \
	(Section~\ref{subsec:LocationScatter});
\item \ $Q = \what{P \!\ominus\! P} = \binom{n}{2}^{-1}
		\sum_{1 \le i < j \le n} \delta_{X_i - X_j}$ \
	(Section~\ref{subsec:SymmetrizedM}).
\end{itemize}

\section{Analytical properties of $L(\cdot,Q)$}
\label{sec:analysis}

As shown in D\"umbgen et al.~\cite{Duembgen_etal_2015}, the functionals $L_0(\cdot,Q)$ and $L_\rho(\cdot,Q)$ are smooth, strictly convex and coercive in a certain sense. To make this precise, we utilize the matrix-valued exponential function: For $A \in \Rqq$ let
\[
	\exp(A) \ := \ \sum_{k=0}^\infty \frac{1}{k!} \, A^k .
\]
In case of $A = A^\top$ we may write $A = U \diag(\lambda) U^\top$ with an orthogonal matrix $U \in \Rqq$ and some vector $\lambda = (\lambda_i)_{i=1}^q \in \R^q$. Then
\[
	\exp(A) \ = \ U \diag(\exp(\lambda)) U^\top
\]
with $\exp(\lambda) := \bigl( \exp(\lambda_i) \bigr)_{i=1}^q$. Moreover,
\[
	\log \det(\exp(A)) \ = \ \tr(A) .
\]
If $A \in \Rqqsympd$, i.e.\ $\lambda \in (0,\infty)^q$, then $A = \exp(\log(A))$ with
\[
	\log(A) \ := \ U \diag(\log(\lambda)) U^\top
\]
and $\log(\lambda) := (\log \lambda_i)_{i=1}^q$.

By means of the matrix-valued exponential function and logarithm, we can describe the behavior of $L(\cdot,Q)$ in a neighborhood of any matrix $\Sigma \in \Rqqsympd$ quite elegantly. Instead of considering additive perturbations $\Sigma + A$ with $A \in \Rqqsym$, we write $\Sigma = BB^\top$ for some nonsingular matrix $B \in \Rqq$, for instance $B = \Sigma^{1/2}$, and consider multiplicative perturbations $B \exp(A) B^\top$. Note that
\[
	\bigl\{ B \exp(A) B^\top : A \in \Rqqsym \bigr\}
	\ = \ \Rqqsympd .
\]
In case of $\det(\Sigma) = 1$,
\[
	\bigl\{ B \exp(A) B^\top : A \in \Rqqsym, \tr(A) = 0 \bigr\}
	\ = \ \bigl\{ \Gamma \in \Rqqsympd : \det(\Gamma) = 1 \bigr\} .
\]
Here is a basic expansion of $L \bigl( B\exp(\cdot)B^\top, Q \bigr)$ around $0$:

\begin{Theorem}[\cite{Duembgen_etal_2015}]
\label{thm:convexity}
For a nonsingular matrix $B \in \Rqq$ define $Q_B := \LL(B^{-1}X)$ with $X \sim Q$. Then for $A \in \Rqqsym$,
\begin{align*}
	L \bigl(
		&B \exp(A) B^\top, Q \bigr) - L \bigl( BB^\top, Q \bigr) \\
	&= \ L(\exp(A), Q_B)
		\ = \ G(A,Q_B) + \frac{1}{2} \, H(A,Q_B) + o(\|A\|^2)
\end{align*}
as $A \to 0$, where
\begin{align*}
	G(A,Q_B) \
	&:= \ \bigl\langle A, I_q - \Psi(Q_B) \bigr\rangle , \\
	H(A,Q_B) \
	&:= \ \bigl\langle A^2, \Psi(Q_B) \bigr\rangle
		+ \int \rho''(\|x\|^2) (x^\top Ax)^2 \, Q_B(dx) ,
\end{align*}
and
\[
	\Psi(Q_B) \ := \ \int \rho'(\|x\|^2) \, xx^\top \, Q_B(dx) .
\]
Moreover, $H(A,Q_B)$ is continuous in $B$, and
\[
	H(A,Q_B) \
	\begin{cases}
		\ge \ 0 , \\
		>   \ 0 & \text{in Setting~0, if} \ A \not\in \{s I_q : s \in \R\}, \\
		>   \ 0 & \text{in Setting~1, if} \ A \ne 0.
	\end{cases}
\]
\end{Theorem}

\begin{Remark}
The Taylor expansion in Theorem~\ref{thm:convexity} implies that
\[
	L(B \exp(A) B^\top, Q) \ = \ L(B \exp(0) B^\top, Q) + \langle A, G(Q_B)\rangle + O(\|A\|^2)
\]
as $A \to 0$, where
\[
	G(Q_B) \ := \ I_q - \Psi(Q_B) \ \in \ \Rqqsym .
\]
Hence the matrix $G(Q_B)$ is the gradient of the function $\Rqqsym \ni A \mapsto L \bigl( B \exp(\cdot) B^\top, Q \bigr)$ at $0 \in \Rqqsym$.

Note also that $\Psi(Q_B)$ is positive definite, because otherwise $Q$ would be concentrated on a proper linear subspace of $\R^q$.
\end{Remark}

\begin{Remark}
Note that $L_0(t\Sigma,Q)$ is constant in $t > 0$ for any $\Sigma \in \Rqqsympd$. In other words, for any nonsingular $B \in \Rqq$, $L_0(B \exp(x I_q) B^\top, Q)$ is constant in $x \in \R$. Applying Theorem~\ref{thm:convexity} to $A = x I_q$ yields that
$G(I_q,Q_B) = \tr(G(Q_B)) = 0$ and $H(I_q,Q_B) = 0$ in Setting~0. This explains the constraint $A \not\in \{s I_q : s \in \R\}$ for $H(A,Q_B) > 0$.
\end{Remark}

\begin{Remark}
The second derivative of the function $L \bigl( B \exp(\cdot) B^\top, Q \bigr)$ at $0 \in \Rqqsym$ corresponds to the quadratic form
\[
	\Rqqsym \times \Rqqsym \ni (A',A)
	\ \mapsto \ \bigl\langle A', H(Q_B) A \bigr\rangle
\]
with the self-adjoint linear operator $H(Q_B) : \Rqqsym \to \Rqqsym$ given by
\[
	H(Q_B)A \ := \ 2^{-1} \bigl( \Psi(Q_B)A + A \Psi(Q_B) \bigr)
		+ \int \rho''(\|x\|^2) x^\top Ax \, xx^\top \, Q_B(dx) .
\]
Theorem~\ref{thm:convexity} implies that this operator is positive definite in Setting~1.
In Setting~0,
\begin{align*}
	\Psi(Q_B) \
	&= \ q \int \|x\|^{-2} \, xx^\top \, Q_B(dx) , \\
	H(Q_B)A \
	&= \ 2^{-1} \bigl( \Psi(Q_B)A + A \Psi(Q_B) \bigr)
		- q \int \|x\|^{-4} x^\top Ax \, xx^\top \, Q_B(dx) ,
\end{align*}
and one easily verifies that $H(Q_B) I_q = 0$ and $\tr \bigl( H(Q_B) A \bigr) = 0$ for any $A \in \Rqqsym$. Hence in both settings one may view $H(Q_B)$ as a self-adjoint and positive definite linear operator from the set
\[
	\mathbb{W} \ := \ \begin{cases}
		\bigl\{ A \in \Rqqsym : \tr(A) = 0 \bigr\} & \text{in Setting~0} \\
		\Rqqsym & \text{in Setting~1}
	\end{cases}
\]
onto itself. In particular, $H(Q_B)^{-1}$ stands for the corresponding inverse mapping.
\end{Remark}

An important consequence of Theorem~\ref{thm:convexity} is a convexity property of $L(\cdot,Q)$:

\begin{Corollary}
\label{cor:convexity}
For any nonsingular $B \in \Rqq$ and $A \in \Rqqsym$, the mapping
\[
	t \ \mapsto \ L \bigl( B \exp(tA) B^\top, Q \bigr)
\]
is twice continuously differentiable and convex on $\R$. In Setting~0 it is strictly convex if $A \not\in \{s I_q : s \in \R\}$. In Setting~1 it is strictly convex if $A \ne 0$.
\end{Corollary}

\noindent
This corollary implies that $\Sigma = BB^\top$ minimizes $L(\cdot,Q)$ if, and only if, the gradient $G(Q_B)$ equals $0$, i.e.
\begin{equation}
\label{eq:fp1}
	\Psi(Q_B) \ = \ I_q .
\end{equation}
This is equivalent to the fixed-point equation
\begin{equation}
\label{eq:fp2}
	\Sigma \ = \ \int \rho'(x^\top \Sigma^{-1}x) xx^\top \, Q(dx) .
\end{equation}

\section{Algorithms}
\label{sec:algorithms}

\subsection{Fixed-point and gradient algorithms}

The fixed-point equation \eqref{eq:fp2} gives rise to a fixed-point algorithm which has been proposed and used repeatedly, see for instance Huber~\cite{Huber_1981} (Section~8.11), Tyler~\cite{Tyler_1987a} and Kent and Tyler~\cite{Kent_Tyler_1991}. The latter two references provide a rigorous proof of convergence for empirical distributions $Q$, the general case is covered by Dudley et al.~\cite{Dudley_etal_2009}. A basic step works as follows: If $\Sigma \in \Rqqsympd$ is our current candidate for a minimizer of $L(\cdot,Q)$, then we replace it with
\[
	\int \rho'(x^\top \Sigma^{-1} x) xx^\top \, Q(dx) .
\]
When implementing this method it is more convenient to utilize the formulation \eqref{eq:fp1} directly: If $\Sigma = BB^\top$ for some nonsingular matrix $B \in \Rqq$, then
\[
	\int \rho'(x^\top \Sigma^{-1} x) xx^\top \, Q(dx)
	\ = \ B \Psi(Q_B) B^\top .
\]
Now we use some factorization $\Psi(Q_B) = CC^\top$ with nonsingular $C \in \Rqq$ and replace $B$ with $BC$. Replacing $\Sigma$ with $B \Psi(Q_B) B^\top$ yields always an improvement, because
\begin{equation}
\label{eq:PsiQ.better.than.I}
	L(B\Psi(Q_B)B^\top,Q) - L(BB^\top, Q) \ < \ 0
	\quad\text{unless} \ \Psi(Q_B) = I_q ;
\end{equation}
see \cite{Duembgen_etal_2015}. Here is a description of the fixed-point algorithm:

\paragraph{Algorithm FP.}
Choose an arbitrary matrix $\Sigma_0 = B_0^{} B_0^\top$ with nonsingular $B_0 \in \Rqq$, and let $Q_0 := Q_{B_0}$. \ Suppose that after $k \ge 0$ steps we have determined a nonsingular matrix $B_k \in \Rqq$, corresponding to the candidate $\Sigma_k = B_k^{}B_k^\top$ for $\bSigma(Q)$. Writing $Q_k := Q_{B_k}$, we compute
\[
	\Psi_k \ := \ \Psi(Q_k) \ = \ \int \rho'(\|x\|^2) \, xx^\top \, Q_k(dx) .
\]
Then we write $\Psi_k = C_k^{}C_k^\top$ for some nonsingular $C_k \in \Rqq$ and define
\[
	B_{k+1} \ := \ B_k C_k .
\]
This corresponds to the new candidate $\Sigma_{k+1} := B_{k+1}^{}B_{k+1}^\top = B_k^{} \Psi_k^{} B_k^\top$.

This description is similar to the one of Huber~\cite{Huber_1981} (Section~8.11), the main difference being that we don't restrict ourselves to the Cholesky factorization of $\Psi_k$. Indeed in our implementation we use $\Psi_k = C_k^{}C_k^\top$ with $C_k = U_k \diag(\phi_k)^{1/2}$, where $\phi_k \in (0,\infty)^q$ contains the eigenvalues of $\Psi_k$ and $U_k$ is an orthogonal matrix of corresponding eigenvectors. Our starting point is typically
\[
	\Sigma_0 \ := \ \int xx^\top \, Q(dx) .
\]
Our stopping criterion for Algorithm~FP is that $\|I_q - \Psi_k\| = \|1_q - \phi_k\| < \delta$ for some given small number $\delta > 0$, where $1_q := (1,1,\ldots,1)^\top \in \R^q$.

An important fact is that under the conditions of Theorem~\ref{thm:existence} the sequence $(\Sigma_k)_{k=0}^\infty$ converges to a minimizer of $L(\cdot,Q)$, no matter which starting point $\Sigma_0$ has been chosen; see also Theorem~\ref{thm:convergence} later.

One may view the fixed-point algorithm as an approximate gradient method with constant stepsize one: Note that with the gradient $G_k := G(Q_k)$ of $L(B_k^{} \exp(\cdot) B_k^\top, Q)$ at $0 \in \Rqqsym$,
\[
	\Sigma_{k+1}
	\ = \ B_k^{} \Psi_k^{} B_k^\top
	\ = \ B_k^{} (I_q - G_k) B_k^\top
	\ = \ B_k^{} \exp \bigl( - G_k + O(\|G_k\|^2) \bigr) B_k^\top .
\]
In the present context an exact gradient method with constant step size one would mean to replace $\Sigma_k$ with $B_k^{} \exp(- G_k) B_k^\top$.

\paragraph{Suboptimality of Algorithm FP.}
As shown later, the steps performed in Algorithm~FP are clearly suboptimal, at least when $\Sigma_k$ is already close to the limit $\bs{\Sigma}(Q)$. To understand this thoroughly and to devise improvements we first provide a corollary to Theorem~\ref{thm:convexity}:

\begin{Corollary}
\label{cor:expansion}
Let $\Sigma = BB^\top$ for a nonsingular matrix $B \in \Rqq$. Further let $Q_* := Q_{\bs{\Sigma}(Q)^{1/2}}$. If we write $B = \Sigma^{1/2} V$ with an orthogonal matrix $V \in \Rqq$, then for any $A \in \Rqqsym$,
\begin{align*}
	L(B \exp(A) B^\top, Q) - L(BB^\top, Q) \
	&= \ L(\exp(A),Q_B) \\
	&= \ G(A,Q_B) + \frac{1}{2}\, H(A,Q_B)
		+ r(B,A)\|A\|^2 \\
	&= \ G(A,Q_B) + \frac{1}{2}\, H(V^\top AV, Q_*)
		+ r_*(B,A)\|A\|^2 ,
\end{align*}
where
\[
	|r(B,A)| + |r_*(B,A)| \ \to \ 0
	\quad\text{as} \ BB^\top \to \bs{\Sigma}(Q) \ \text{and} \ A \to 0 .
\]
Moreover,
\[
	H(V^\top AV, Q_*)
	\ = \ \|A\|^2 + \int \rho''(\|x\|^2) (x^\top V^\top AV x)^2 \, Q_*(dx) .
\]
\end{Corollary}

Now let us apply this corollary to Algorithm~FP. We write $B_k = \Sigma_k^{1/2} V_k$ for some orthogonal matrix $V_k \in \Rqq$. If we fix an arbitrary constant $K > 1$, then uniformly in $A \in \Rqqsym$ with $\|A\| \le K \|G_k\|$,
\begin{align*}
	L(B_k^{} \exp(A) B_k^\top, Q) - L(B_k^{}B_k^\top, Q) \
	&= \ L(\exp(A), Q_k) \\
	&= \ \langle A, G_k\rangle + \frac{1}{2}\, H(V_k^\top A V_k^{}, Q_*)
		+ r_*(B_k,A) \|A\|^2 \\
	&= \ \langle A, G_k\rangle + \frac{1}{2}\, H(V_k^\top A V_k^{}, Q_*)
		+ o(\|G_k\|^2) .
\end{align*}
In particular, if we choose $A = - t_k G_k$ with a bounded sequence $(t_k)_k$ in $\R$,
\[
	L(\exp(- t_kG_k), Q_k)
	\ = \ \|G_k\|^2 \Bigl( - t_k
		+ \frac{t_k^2}{2} \, \frac{H(V_k^\top G_k^{} V_k^{}, Q_*)}{\|G_k\|^2} + o(1) \Bigr) .
\]
Consequently, an approximately optimal choice of $t_k$ would be a minimizer of the right hand side without the term $o(1)$, i.e.
\begin{align*}
	t_k^* \
	&= \ \frac{\|G_k\|^2}{H(V_k^\top G_k^{} V_k^{}, Q_*)} \\
	&= \ \Bigl( 1 + \int \rho''(\|x\|^2)
			\frac{(x^\top V_k^\top G_k^{} V_k^{} x)^2}{\|G_k\|^2} \, Q_*(dx) \Bigr)^{-1} \\
	&\in \ \biggl[
		\Bigl( 1 - \min_{A \in \mathbb{W} : \|A\| = 1}
		\int |\rho''|(\|x\|^2) (x^\top A x)^2 \, Q_*(dx) \Bigr)^{-1} ,
		\lambda_{\rm min} \bigl( H(Q_*) \bigr)_{}^{-1} \biggr] .
\end{align*}
The upper bound involves the minimal eigenvalue of the symmetric operator $H(Q_*) : \mathbb{W} \to \mathbb{W}$. The lower bound follows from $\rho'' \le 0$ and is typically strictly larger than $1$, for instance if $\rho = \rho_{\nu,q}$ as defined in \eqref{eq:rho_nu} or \eqref{eq:rho_0}. Hence the steps performed during the fixed-point algorithm tend to be too short!

\paragraph{Algorithm G.}
One could easily fix this deficiency as follows: As a proxy for $t_k^*$, which involves the unknown quadratic form $H(\cdot,Q_*)$, we compute in the $k$-th iteration the number
\[
	t_k
	\ = \ \frac{\|G_k\|^2}{H(G_k, Q_k)}
	\ = \ t_k^* \, (1 + o(1)) .
\]
The latter equality follows from Corollary~\ref{cor:expansion}. Indeed, the latter corollary implies that we obtain $L(\exp(- t_k G_k), Q_k) = - \|G_k\|^4 / (2H(G_k, Q_k)) (1 + o(1)) \le - \|G_k\|^2/2 (1 + o(1))$. Thus we check whether
\begin{equation}
\label{eq:Check.G.step}
	L \bigl( \exp(- t_k G_k), Q_k \bigr)
	\ \le \ - \|G_k\|^2/4 .
\end{equation}
If yes, we replace $B_k$ with $B_{k+1} = B_k C_k$, where $C_k^{}C_k^\top = \exp(- t_k G_k)$. Otherwise we perform a usual fixed-point step as described before. The number $4$ in \eqref{eq:Check.G.step} could be replaced with any number $c > 2$.

Implementing this gradient method yielded already a substantial reduction of computation time. This approach of improving a fixed-point algorithm by means of variable step lengths is also used by Redner and Walker~\cite{Redner_Walker_1984} in the context of maximum-likelihood estimation for mixture models. But in view of Theorem~\ref{thm:convexity} it is certainly tempting to try a Newton-Raphson procedure.

\subsection{(Partial) Newton-Raphson procedures}

Suppose that our current candidate for $\bs{\Sigma}(Q)$ is $\Sigma = BB^\top$. In view of Corollary~\ref{cor:expansion} we should replace $\Sigma$ with
\[
	\tilde{\Sigma} \ = \ B \exp \bigl( - H(Q_B)^{-1} G(Q_B) \bigr) B^\top ,
\]
because $H(Q_B)^{-1} G(Q_B)$ is the unique minimizer of
\[
	\mathbb{W} \ni A \ \mapsto \ G(A,Q_B) + \frac{1}{2} H(A, Q_B)
	\ = \ \langle A, G(Q_B)\rangle + \frac{1}{2} \langle A, H(Q_B) A\rangle .	
\]
A problem with this promising update $\tilde{\Sigma}$ is that the computation of the inverse operator $H(Q_B)^{-1}$ may be too computer- or memory-intensive. Indeed, we implemented a full Newton-Raphson algorithm, and it required only very few iterations, as expected. But the running time was even longer than with Algorithm~FP, because the computation and inversion of $H(Q_B)$, which may be represented by a symmetric matrix in $\R^{\dim(\mathbb{W})\times\dim(\mathbb{W})}$, was too time-consuming. Note that $\dim(\mathbb{W})$ equals $q(q+1)/2 - 1$ in Setting~0 and $q(q+1)/2$ in Setting~1.

These difficulties with a full Newton-Raphson procedure have been noticed already by Huber~\cite{Huber_1981} (Section~8.11). Some authors have tried alternative approaches such as conjugate gradient methods or quasi Newton methods in which the operator $H(Q_B)$ is replaced with a surrogate which is easier to compute and invert; see for instance Huber~\cite{Huber_1977a}. According to \cite{Huber_1981}, none of these attempts was overall convincing.

A partial Newton-Raphson approach turned out to be quite successful. This means that instead of considering arbitrary multiplicative perturbations $B \exp(A) B^\top$ of a current candidate $\Sigma = BB^\top$, we restrict $A$ to a particular $q$-dimensional subspace of $\Rqqsym$ depending on $B$. Precisely, consider the matrix $\Psi(Q_B) \in \Rqqsympd$ and its spectral decomposition,
\[
	\Psi(Q_B) \ = \ U \diag(\phi) U^\top
\]
with an orthogonal matrix $U \in \Rqq$ whose columns are eigenvectors of $\Psi(Q_B)$ and a vector $\phi \in (0,\infty)^q$ containing the corresponding eigenvalues. Now we consider only perturbations $\Sigma = B \exp(A) B^\top$ with $A = U \diag(a) U^\top$, $a \in \R^q$. Since $\exp(U\diag(a)U^\top) = U \exp(\diag(a)) U^\top$, this leads to the functional
\[
	\R^q \ni a
	\ \mapsto \ L \bigl( B \, U \exp(\diag(a)) U^\top B^\top, Q) - L(BB^\top, Q) .
\]
Now the Taylor expansion in Theorem~\ref{thm:convexity} may be rewritten as follows:
\begin{align*}
	L \bigl( B \, U &\exp(\diag(a)) U^\top B^\top, Q) - L(BB^\top, Q) \\
	&= \ L \bigl( \exp(\diag(a)), Q_{BU} \bigr)
		\ = \ \tilde{G}(Q_{BU})^\top a + \frac{1}{2} \, a^\top \tilde{H}(Q_{BU}) a
		+ o(\|a\|^2) ,
\end{align*}
where
\begin{align*}
	\tilde{G}(Q_{BU}) \
	&:= \ 1_q - \int \rho'(\|x\|^2) s(x) \, Q_{BU}(dx)
		\ = \ 1_q - \phi
		\ \in \ \R^q , \\
	\tilde{H}(Q_{BU}) \
	&:= \ \diag(\phi) + \int \rho''(\|x\|^2) s(x) s(x)^\top \, Q_{BU}(dx)
		\ \in \ \Rqqsym
\end{align*}
with $1_q = (1)_{j=1}^q$ and
\[
	s(x) \ := \ (x_j^2)_{j=1}^q
	\quad\text{for} \ x = (x_j)_{j=1}^q \in \R^q .
\]
In Setting~1, $\tilde{H}(Q_{BU})$ is a positive definite matrix, and
\[
	\argmin_{a \in \R^q} \bigl( \tilde{G}(Q_{BU})^\top a
		+ \frac{1}{2} \, a^\top \tilde{H}(Q_{BU}) a \bigr)
	\ = \ - \tilde{H}(Q_{BU})^{-1} \tilde{G}(Q_{BU}) .
\]
In Setting~0, the matrix $\tilde{H}(Q_{BU})$ satisfies $\tilde{H}(Q_{BU}) 1_q = 0$ and $a^\top \tilde{H}(Q_{BU}) a > 0$ whenever $a \ne 0$, $1_q^\top a = 0$. Moreover, $1_q^\top \tilde{G}(Q_{BU}) = 0$. Thus we may write
\[
	\argmin_{a \in \R^q} \bigl( \tilde{G}(Q_{BU})^\top a
		+ \frac{1}{2} \, a^\top \tilde{H}(Q_{BU}) a \bigr)
	\ = \ - (\tilde{H}(Q_{BU}) + c \, 1_q^{}1_q^\top)^{-1} \tilde{G}(Q_{BU})
\]
for any constant $c > 0$.

\paragraph{Algorithm PN.}
Choose an arbitrary matrix $\Sigma_0 = B_0^{} B_0^\top$ with nonsingular $B_0 \in \Rqq$, and let $Q_0 := Q_{B_0}$.\\
Suppose that for some integer $k \ge 0$ we have already determined a nonsingular matrix $B_k \in \Rqq$. Writing $Q_k := Q_{B_k}$, we compute
\[
	\Psi_k \ := \ \Psi(Q_k) \ = \ \int \rho'(\|x\|^2) \, xx^\top \, Q_k(dx) .
\]
Then we write $\Psi_k = U_k^{} \diag(\phi_k) U_k^\top$ with an orthogonal matrix $U_k \in \Rqq$ and a vector $\phi_k \in (0,\infty)^q$. Next we define
\[
	\tilde{Q}_k \ := \ (Q_k)_{U_k} \ = \ Q_{B_kU_k}
\]
and
\[
	a_k \ := \ \begin{cases}
		- \tilde{H}(\tilde{Q}_k)^{-1} \tilde{G}(\tilde{Q}_k)
		& \text{in Setting~1} , \\
		- \bigl( \tilde{H}(\tilde{Q}_k) + c\,1_q^{}1_q^\top \bigr)^{-1} \tilde{G}(\tilde{Q}_k)
		& \text{in Setting~0} .
	\end{cases}
\]
We expect that replacing $B_k$ with $B_k \exp(\diag(a_k/2))$ results in a change of $L(\cdot,Q)$ of about $a_k^\top \tilde{G}(\tilde{Q}_k)/2 < 0$. Now we check whether
\begin{equation}
\label{eq:Check.PN.step}
	L \bigl( \exp(\diag(a_k)), \tilde{Q}_k \bigr)
	\ \le \ a_k^\top \tilde{G}(\tilde{Q}_k)/4 .
\end{equation}
If yes, we define
\[
	B_{k+1} \ := \ B_k U_k \exp(\diag(a_k/2))
\]
which corresponds to the new candidate $\Sigma_{k+1} := B_{k+1}^{} B_{k+1}^\top = B_k^{} \exp(\diag(a_k^{})) B_k^\top$. If \eqref{eq:Check.PN.step} is violated we just perform a step of the fixed-point algorithm and set $B_{k+1} := B_k U_k \diag(\phi_k)^{1/2}$, i.e.\ our new candidate is $\Sigma_{k+1} := B_{k+1}^{} B_{k+1}^\top = B_k^{} \diag(\phi_k^{}) B_k^\top$. Again, the number $4$ in \eqref{eq:Check.PN.step} could be replaced by any number $c > 2$.

The new Algorithm~PN is also guaranteed to converge to a minimizer of $L(\cdot,Q)$:

\begin{Theorem}
\label{thm:convergence}
For any starting point $\Sigma_0 \in \Rqqsympd$ and in both Settings~0 and 1, Algorithm~FP as well as Algorithm~PN yield a sequence $(\Sigma_k)_k$ converging to a minimizer $\Sigma_*$ of $L(\cdot,Q)$.
\end{Theorem}

For general distributions $Q$ it is difficult to compare Algorithms~FP and PN explicitly. Recall that in Algorithm~PN we restrict our attention to a particular subspace of $\Rqqsympd$. The following lemma implies that at least in case of an (approximately) elliptically symmetric distribution $Q$ this subspace is (almost) the right one to look in for better candidates.

\begin{Lemma}
\label{lem:PN.is.right}
Suppose that $Q$ is elliptically symmetric with center $0$ and scatter matrix $\Sigma_o \in \Rqqsympd$. Then $\bs{\Sigma}(Q) = \kappa \Sigma_o$ for some $\kappa > 0$. Moreover, for any $\Sigma = BB^\top$ with nonsingular $B \in \Rqq$ and any spectral decomposition $\Psi(Q_B) = U \diag(\phi) U^\top$,
\[
	\bs{\Sigma}(Q) \ = \ B U \exp(\diag(a)) U^\top B^\top
\]
for a vector $a \in \R^q$ containing the log-eigenvalues of $\Sigma^{-1} \bs{\Sigma}(Q)$.
\end{Lemma}

At this point we should mention that for ``well-behaved'' distributions $Q$ in high dimension $q$, algorithm FP can be rather efficient, because the standardized distribution $Q_* = Q_{\bSigma(Q)^{1/2}}$ satisfies
\[
	H(A,Q_*) \ \approx \ \|A\|^2
\]
for $A \in \mathbb{W}$. For instance in Setting~0, if $Q_*$ is spherically symmetric around $0$,
\[
	H(A,Q_*) \ = \ \frac{q}{q+2} \|A\|^2
\]
for all $A \in \mathbb{W}$. Hence, if $\Sigma = BB^\top$ is already close to $\bSigma(Q)$, the Newton step would be to replace $\Sigma$ with
\[
	\Sigma_{\rm new} \ \approx \ B \exp( - (1 + 2/q) G(Q_B)) B^\top ,
\]
and for high dimension $q$ this is similar to $B \exp( - G(Q_B)) B^\top \approx B\Psi(Q_B)B^\top$. Indeed our numerical experiments show that Algorithm~PN is particularly useful in situations where $Q$ is ``problematic'', e.g.\ an empirical distribution of a sample with strong outliers.

\subsection{Explicit pseudo-code}

\paragraph{Standard $M$-estimators.}
Suppose that $Q = \sum_{i=1}^n w_i \delta_{x_i}$ with a certain weight vector $\bs{w} = (w_i)_{i=1}^n$ in $(0,1)^n$ such that $\sum_{i=1}^n w_i = 1$ and a data matrix $\bs{X} = [x_1, x_2, \ldots, x_n]^\top \in \R^{n\times q}$. Then our Algorithm~PN may be formulated as in Table~\ref{tab:PN}.

\begin{table}
\[
\begin{array}{|l|}
	\hline
	\Sigma \leftarrow \mathbf{AlgorithmPN}(\bs{X},\bs{w},\delta) \\[1ex]
	B \leftarrow \bigl( \sum_{i=1}^n w_i^{} x_i^{}x_i^\top \bigr)_{}^{1/2}\\
	\bs{Y} \leftarrow \bs{X} B^{-1} \\
	\Psi \leftarrow \sum_{i=1}^n w_i \rho'(\|y_i\|^2) \, y_i^{}y_i^\top\\
	(U,\phi) \leftarrow \mathrm{Eigen}(\Psi)\\[1ex]
	\mathbf{while} \ \|1_q - \phi\| > \delta \ \mathbf{do}\\
	\strut\qquad
		B \leftarrow B U\\
	\strut\qquad
		\bs{Y} \leftarrow \bs{Y} U\\
	\strut\qquad
		\tilde{H} \leftarrow \diag(\phi)
			+ \sum_{i=1}^n w_i \rho''(\|y_i\|^2) s(y_i) s(y_i)^\top
			\ (+ \ c \, 1_q^{}1_q^\top \ \text{in Setting~0})\\
	\strut\qquad
		a \leftarrow \tilde{H}^{-1} (\phi - 1_q)\\
	\strut\qquad
		\bs{Z} \leftarrow \bs{Y} \exp(- \diag(a)/2)\\
	\strut\qquad
		DL \leftarrow \sum_{i=1}^n w_i \bigl( \rho(\|z_i\|^2) - \rho(\|y_i\|^2) \bigr)
			+ \sum_{j=1}^q a_j\\
	\strut\qquad
		DL_0 \leftarrow a^\top (1_q - \phi)/4\\
	\strut\qquad
		\mathbf{if} \ DL \le DL_0 \ \mathbf{then}\\
	\strut\qquad\qquad
			B \leftarrow B \exp(\diag(a)/2)\\
	\strut\qquad\qquad
			\bs{Y} \leftarrow \bs{Z}\\
	\strut\qquad
		\mathbf{else}\\
	\strut\qquad\qquad
			B \leftarrow B \diag(\phi)^{1/2}\\
	\strut\qquad\qquad
			\bs{Y} \leftarrow \bs{Y} \diag(\phi)^{-1/2}\\
	\strut\qquad
		\mathbf{end~if}\\
	\strut\qquad
		\Psi \leftarrow \sum_{i=1}^n w_i \rho'(\|y_i\|^2) \, y_i^{}y_i^\top\\
	\strut\qquad
		(U,\phi) \leftarrow \mathrm{Eigen}(\Psi)\\
	\mathbf{end~while}\\[1ex]
	\Sigma \leftarrow BB^\top\\
	\mathbf{return} \ \Sigma\\
	\hline	
	\end{array}
\]
\caption{Pseudo-code for the $M$-estimator.}
\label{tab:PN}
\end{table}

\paragraph{Symmetrized $M$-estimators.}
Suppose that
\[
	Q \ = \ \binom{n}{2}^{-1} \sum_{1 \le i < j \le n} \delta_{x_i - x_j}^{}
\]
for a certain data matrix $\bs{X} = [x_1, x_2, \ldots, x_n]^\top \in \R^{n\times q}$. In principle one could utilize the algorithm just described with $N = \binom{n}{2}$ in place of $n$ and $\bs{X}$ replaced by a data matrix $\tilde{\bs{X}}$ containing all $N$ differences $x_i - x_j$. For large $n$, however, this may require too much computer memory, and one should avoid the explicit storage of such a large data matrix $\tilde{\bs{X}}$.

It turned out that the computation time can be reduced substantially if we first compute the $M$-estimator $\bs{\Sigma}(\tilde{Q})$ for the surrogate distribution
\[
	\tilde{Q} \ := \ \frac{1}{n} \sum_{i=1}^n \delta_{x_{\pi(i)} - x_{\pi(i+1)}}
\]
with a randomly chosen permutation $\pi$ of $\{1,2,\ldots,n\}$ and $\pi(n+1) := \pi(1)$. Then we use this estimator $\bs{\Sigma}(\tilde{Q})$ as a starting parameter $\Sigma_0$ in Algorithm~PN.

Table~\ref{tab:PN_symm} contains pseudo-code for the computation of the symmetrized $M$-estimator without using a large data matrix $\tilde{\bs{X}}$. Instead it utilizes auxiliary programs to compute the following objects:
\begin{align*}
	\mathrm{RPermute}(n) \
	&\rightarrow \ \text{a random permutation of} \ \{1,2,\ldots,n\} , \\
	\mathrm{Psi}(\bs{X}) \
	&\rightarrow \ \frac{1}{N} \sum_{1 \le i < j \le n}
		\rho'(\|x_i - x_j\|^2) (x_i - x_j)(x_i - x_j)^\top , \\
	\mathrm{H}(\phi,\bs{X}) \
	&\rightarrow \ \diag(\phi) + \frac{1}{N} \sum_{1 \le i < j \le n}
		\rho''(\|x_i - x_j\|^2) s(x_i - x_j) s(x_i - x_j)^\top , \\
	\mathrm{DL}(\bs{X},\bs{Y},a) \
	&\rightarrow \ \frac{1}{N} \sum_{1 \le i < j \le n}
		\bigl[ \rho(\|y_i - y_j\|^2) - \rho(\|x_i - x_j\|^2) \bigr]
		+ \sum_{k=1}^q a_k .
\end{align*}

\begin{table}
\[
\begin{array}{|l|}
	\hline
	\Sigma \leftarrow \mathbf{AlgorithmPN.symm}(\bs{X},\delta) \\[1ex]
	\pi \leftarrow \mathrm{RPermute}(n)\\
	\bs{X}^0 \leftarrow [x_{\pi(1)} - x_{\pi(2)}, x_{\pi(2)} - x_{\pi(3)},
		\ldots, x_{\pi(n)} - x_{\pi(1)}]_{}^\top\\
	B \leftarrow \mathbf{AlgorithmPN}(\bs{X}^0,(1/n)_{i=1}^n,\delta)_{}^{1/2}\\
	\bs{Y} \leftarrow \bs{X} B^{-1} \\
	\Psi \leftarrow \mathrm{Psi}(\bs{Y})\\
	(U,\phi) \leftarrow \mathrm{Eigen}(\Psi)\\[1ex]
	\mathbf{while} \ \|1_q - \phi\| > \delta \ \mathbf{do}\\
	\strut\qquad
		B \leftarrow B U\\
	\strut\qquad
		\bs{Y} \leftarrow \bs{Y} U\\
	\strut\qquad
		\tilde{H} \leftarrow \mathrm{H}(\phi,\bs{Y}) \ \
		(+ \ c \, 1_q^{}1_q^\top \ \text{in Setting~0}) \\
	\strut\qquad
		a \leftarrow \tilde{H}^{-1} (\phi - 1_q)\\
	\strut\qquad
		\bs{Z} \leftarrow \bs{Y} \exp(- \diag(a)/2)\\
	\strut\qquad
		DL \leftarrow \mathrm{DL}(\bs{Y},\bs{Z},a)\\
	\strut\qquad
		DL_0 \leftarrow a^\top (1_q - \phi)/4\\
	\strut\qquad
		\mathbf{if} \ DL \le DL_0 \ \mathbf{then}\\
	\strut\qquad\qquad
			B \leftarrow B \exp(\diag(a)/2)\\
	\strut\qquad\qquad
			\bs{Y} \leftarrow \bs{Z}\\
	\strut\qquad
		\mathbf{else}\\
	\strut\qquad\qquad
			B \leftarrow B \diag(\phi)^{1/2}\\
	\strut\qquad\qquad
			\bs{Y} \leftarrow \bs{Y} \diag(\phi)^{-1/2}\\
	\strut\qquad
		\mathbf{end~if}\\
	\strut\qquad
		\Psi \leftarrow \mathrm{Psi}(\bs{Y})\\
	\strut\qquad
		(U,\phi) \leftarrow \mathrm{Eigen}(\Psi)\\
	\mathbf{end~while}\\[1ex]
	\Sigma \leftarrow BB^\top\\
	\mathbf{return} \ \Sigma\\
	\hline	
	\end{array}
\]
\caption{Pseudo-code for the symmetrized $M$-estimator.}
\label{tab:PN_symm}
\end{table}

\section{Numerical examples and comparisons}
\label{sec:Examples}

In most of our simulation experiments we simulated data matrices $\bs{X} = [X_1, X_2, \ldots, X_n]^\top$ with independent rows $X_i = (X_{ij})_{j=1}^q$ having either standard Gaussian or standard Cauchy distribution on $\R^q$. In the latter case, $(X_{ij})_{j=1}^q$ is distributed as $(Z_j/Z_0)_{j=1}^q$ with independent random variables $Z_0, Z_1, \ldots, Z_q \sim \NN(0,1)$. In all experiments, iterations were stopped when the gradient $G_k = G(Q_k)$ of our target function satisfies $\|G_k\| \le 10^{-7}$, and the number of Monte Carlo simulations for each setting was $500$.

The first three experiments were run on a MacBook Pro (2GHz Intel(R) Core i7, 16GB), the fourth experiment on a Windows server (two Intel(R) Xeon(R) CPU R5 2440 with 2.40GHz and 64GB). We used R 3.1.2 \cite{R_2013}.

\paragraph{Comparisons in scatter-only settings.}
To compare the three algorithms FP, G and PN, we first implemented them in pure R code. Table~\ref{tab:Comparisons1} contains the mean number of iterations and the mean computing times for the scatter estimator $\bs{\Sigma}(\what{P})$ with $\rho = \rho_{1,q}$ based on a data matrix $\bs{X} \in \R^{500\times q}$, $q = 5, 10, 20$. The table entries are the mean iteration numbers and mean computations times in milliseconds [ms]. In brackets the corresponding inter quartile ranges are recorded as well. The relative efficiencies are the ratios of the mean computation times. Algorithm~G is already more efficient than Algorithm~FP, but obviously Algorithm~PN is substantially faster than the other two, and this advantage grows with the dimension $q$. Note also that computation costs are higher for Cauchy data than for Gaussian data.

\begin{table}
\[
	\begin{array}{|l|c|c|c||c|c|c|}
	\cline{2-7}
	\multicolumn{1}{l}{}
	& \multicolumn{3}{|l||}{\text{Gaussian data}}
	& \multicolumn{3}{|l|}{\text{Cauchy data}} \\
	\hline
	\text{Algorithm}
	& \text{FP} & \text{G} & \text{PN}
	& \text{FP} & \text{G} & \text{PN} \\
	\hline
	\multicolumn{7}{c}{} \\[-1.5ex]
	\cline{1-1}
	\bs{q = 5} \\
	\hline
	\text{Iterations}
	&  83.9 \ (2) & 31.2 \  (4) & 5.1 \ (0)
	& 116.4 \ (3) & 45.5 \ (14) & 8.5 \ (1) \\
	\hline
	\text{Time [ms]}
	& 13.5 \ (0.5) & 11.4 \ (1.8) & 1.8 \ (0.3)
	& 18.5 \ (1.0) & 16.8 \ (5.3) & 2.8 \ (0.5) \\
	\hline
	\text{Relative}
	& \text{FP}
	       & 1.18 & \bs{7.71}
	& \text{FP}
	       & 1.10 & \bs{6.53} \\
	\text{efficiency}
	&      & \text{G}
	              & \bs{6.51}
	&      & \text{G}
	              & \bs{5.95} \\
	\hline
	\multicolumn{7}{c}{} \\[-1.5ex]
	\cline{1-1}
	\bs{q = 10} \\
	\hline
	\text{Iterations}
	& 141.6 \ (1) & 46.0 \  (6) & 6.0 \ (0)
	& 189.4 \ (3) & 69.4 \ (30) & 9.3 \ (1) \\
	\hline
	\text{Time [ms]}
	& 41.9 \ (1.0) & 25.0 \  (2.8) & 3.1 \ (0.3)
	& 56.2 \ (2.3) & 37.1 \ (16.0) & 5.0 \ (1.0) \\
	\hline
	\text{Relative}
	& \text{FP}
	       & 1.68 & \bs{13.37}
	& \text{FP}
	       & 1.51 & \bs{11.19} \\
	\text{efficiency}
	&      & \text{G}
	              & \bs{7.97}
	&      & \text{G}
	              & \bs{7.40} \\
	\hline
	\multicolumn{7}{c}{} \\[-1.5ex]
	\cline{1-1}
	\bs{q = 20} \\
	\hline
	\text{Iterations}
	& 252.2 \ (2) & 119.2 \  (6) &  6.0 \ (0)
	& 332.2 \ (4) & 103.7 \ (43) & 10.6 \ (1) \\
	\hline
	\text{Time [ms]}
	& 176.2 \ (4.8) & 120.2 \  (7.8) &  6.9 \ (0.3)
	& 230.1 \ (4.8) & 104.4 \ (43.4) & 12.4 \ (1.3) \\
	\hline
	\text{Relative}
	& \text{FP}
	       & 1.47 & \bs{25.65}
	& \text{FP}
	       & 2.20 & \bs{18.54} \\
	\text{efficiency}
	&      & \text{G}
	              & \bs{17.49}
	&      & \text{G}
	              & \bs{8.41} \\
	\hline
	\end{array}
\]
\caption{Computation costs and relative efficiencies in scatter-only settings ($n = 500$, $\nu = 1$).}
\label{tab:Comparisons1}
\end{table}

\paragraph{Comparisons in location-scatter settings.}
Now we consider the empirical distribution $\what{P}$ of the rows of $\bs{X}$ and for given $\nu \ge 1$ the minimizer $\bigl( \bmu_\nu(\what{P}),\bSigma_\nu(\what{P}) \bigr)$ of
\[
	L_\nu(\mu,\Sigma,\what{P})
	\ := \ L_\nu(\Gamma(\mu,\Sigma),\what{Q})
\]
over all $(\mu,\Sigma) \in \R^q \times \Rqqsympd$. Here $\Gamma(\mu,\Sigma) \in \R^{(q+1)\times(q+1)}_{{\rm sym},>0}$ is defined as in \eqref{eq:augmentation}, $\what{Q}$ stands for the empirical distribution of the augmented data points $[X_i^\top,1]^\top \in \R^{q+1}$, $1 \le i \le n$, and
\[
	L_\nu(\Gamma,\what{Q})
	\ := \ \int \bigl[ \rho_{\nu-1,q+1}(y^\top \Gamma^{-1}y)
		- \rho_{\nu-1,q+1}(y^\top y) \bigr] \, \what{Q}(dy) + \log\det(\Gamma)
\]
for arbitrary $\Gamma \in \R^{(q+1)\times(q+1)}_{{\rm sym},>0}$.

In principle, we may apply any of the three algorithms FP, G and PN to the empirical distribution $\what{Q}$ to compute a minimizer $\what{\Gamma}$ of $L_\nu(\cdot,\what{Q})$. In case of $\nu > 1$ this minimizer satisfies automatically $\what{\Gamma}_{q+1,q+1} = 1$, so $\what{\Gamma} = \Gamma \bigl( \bmu_\nu(\what{P}),\bSigma_\nu(\what{P}) \bigr)$. In case of $\nu = 1$, $\what{\Gamma}$ equals $\Gamma \bigl( \bmu_\nu(\what{P}),\bSigma_\nu(\what{P}) \bigr)$ times $\what{\Gamma}_{q+1,q+1}$.

In addition we implemented a variant FP$_3$ of FP proposed by Arslan et al.~\cite{Arslan_etal_1995}. Suppose that $(\mu_k,B_k^{}B_k^\top)$ with nonsingular $B_k \in \Rqq$ is a current candidate for $\bigl( \bmu_\nu(\what{P}),\bSigma_\nu(\what{P}) \bigr)$. Let $\what{Q}_k$ denote the empirical distribution of the standardized data points $B_k^{-1}(X_i - \mu_k)$, $1 \le i \le n$, augmented by an additional component $1$, and define
\[
	\Psi_k \ := \ \int \rho_{\nu-1,q+1}'(y^\top y) yy^\top \, \what{Q}_k(dy) .
\]
Recall that $(\mu_k,B_k^{}B_k^\top)$ equals $\bigl( \bmu_\nu(\what{P}),\bSigma_\nu(\what{P}) \bigr)$ if, and only if, $\Psi_k = I_{q+1}$. Now we write $\Psi_k = \lambda_k \Gamma(\delta_k, C_k^{}C_k^\top)$ for some $\lambda_k > 0$, $\delta_k \in \R^k$ and a nonsingular matrix $C_k \in \Rqq$. Then the next candidate for $\bigl( \bmu_\nu(\what{P}),\bSigma_\nu(\what{P}) \bigr)$ equals
$(\mu_{k+1}, B_{k+1}^{}B_{k+1}^\top)$ with
\[
	\mu_{k+1} \ := \ \mu_k + B_k \delta_k ,
	\quad
	B_{k+1} \ := \ B_k C_k .
\]
To provide a fair comparison, we used the same stopping criterion as for the other algorithms, that means, we considered the norm of $I_{q+1} - \Psi_k$.

For $n = 100$ and $q = 10$ we simulated data matrices $\bs{X} \in \R^{n\times q}$ with independent entries
\[
	X_{ij} \ \sim \ \begin{cases}
		\NN(\delta,1) & \text{if} \ i \le n/10 \ \text{and} \ j = 1 , \\
		\NN(0,1)      & \text{else} ,
	\end{cases}
\]
where $\delta \ge 0$ is a certain parameter quantifying the outlyingness of the $n/10$ first data vectors. The left and right half of Table~\ref{tab:Comparisons2} show the resulting computation costs and times for $\delta = 0, 10, 20$ when $\nu = 1$ and $\nu = 2$, respectively. For $\nu = 1$, algorithm FP is more efficient than FP$_3$. Indeed one can easily verify that the two algorithms are essentially equivalent, the only difference being how they factorize matrices such as $\Psi_k$. For $\delta = 0$, algorithm FP ($\nu = 1$) and algorithm FP$_3$ ($\nu = 2$) are remarkably efficient and even outperform algorithm PN. But for larger values of $\delta$, leading to heterogeneous data sets, PN is clearly the fastest method.

\begin{table}
\[
	\begin{array}{|l|c|c|c||c|c|c|}
	\cline{2-7}
	\multicolumn{1}{l}{}
	& \multicolumn{3}{|l||}{\nu = 1}
	& \multicolumn{3}{|l|}{\nu = 2} \\
	\hline
	\!\text{Algorithm}
	& \text{FP} & \text{FP$_3$} & \text{PN}
	& \text{FP} & \text{FP$_3$} & \text{PN} \\
	\hline
	\multicolumn{6}{c}{} \\[-1.5ex]
	\cline{1-1}
	\bs{\delta = 0} \\
	\hline
	\text{Iterations}
	&  15.1 \ (0) & 15.1 \ (0) & 9.6 \ (1)
	& 152.0 \ (3) & 13.8 \ (1) & 8.9 \ (0) \\
	\hline
	\text{Time [ms]}
	&  2.3 \ (0.2) & 2.7 \ (0.2) & 3.0 \ (0.2)
	& 21.8 \ (0.6) & 2.8 \ (0.3) & 2.9 \ (0.1) \\
	\hline
	\text{Relative}
	& \text{FP}
	       & 0.87 & \bs{0.77}
	& \text{FP}
	       & 7.81 & \bs{7.62} \\
	\text{efficiency}
	&      & \text{FP$_3$}
	              & \bs{0.88}
	&      & \text{FP$_3$}
	              & \bs{0.98} \\
	\hline
	\multicolumn{7}{c}{} \\[-1.5ex]
	\cline{1-1}
	\bs{\delta = 10} \\
	\hline
	\text{Iterations}
	&  27.4 \ (4) & 27.4 \ (4) & 12.3 \ (1)
	& 157.3 \ (3) & 25.8 \ (3) & 11.6 \ (1) \\
	\hline
	\text{Time [ms]}
	&  4.0 \ (0.6) & 4.7 \ (0.7) & 3.7 \ (0.3)
	& 22.3 \ (0.6) & 4.9 \ (0.6) & 3.7 \ (0.3) \\
	\hline
	\text{Relative}
	& \text{FP}
	       & 0.85 & \bs{1.09}
	& \text{FP}
	       & 4.60 & \bs{6.11} \\
	\text{efficiency}
	&      & \text{FP$_3$}
	              & \bs{1.28}
	&      & \text{FP$_3$}
	              & \bs{1.33} \\
	\hline
	\multicolumn{7}{c}{} \\[-1.5ex]
	\cline{1-1}
	\bs{\delta = 20} \\
	\hline
	\text{Iterations}
	&  47.2 \ (6) & 47.2 \ (6) & 17.2 \ (2)
	& 161.4 \ (3) & 42.0 \ (4) & 15.6 \ (1) \\
	\hline
	\text{Time [ms]}
	&  6.6 \ (0.9) & 7.8 \ (1.0) & 5.0 \ (0.5)
	& 23.0 \ (0.6) & 7.9 \ (1.0) & 4.9 \ (0.4) \\
	\hline
	\text{Relative}
	& \text{FP}
	       & 0.84 & \bs{1.31}
	& \text{FP}
	       & 2.93 & \bs{4.66} \\
	\text{efficiency}
	&      & \text{FP$_3$}
	              & \bs{1.56}
	&      & \text{FP$_3$}
	              & \bs{1.59} \\
	\hline
	\end{array}
\]
\caption{Computation costs and relative efficiencies in location-scatter settings ($q = 10$, $n = 100$).}
\label{tab:Comparisons2}
\end{table}

\paragraph{Comparisons for symmetrized scatter estimators, I.}
As mentioned in the introduction, computation time becomes a major issue when computing symmetrized scatter estimators. In the simulation experiments described below we simulated data matrices $\bs{X} \in \R^{n \times q}$ with independent rows following a multivariate standard Gaussian or standard Cauchy distribution on $\R^q$.

\begin{figure}
\centering
\includegraphics[width=0.9\textwidth]{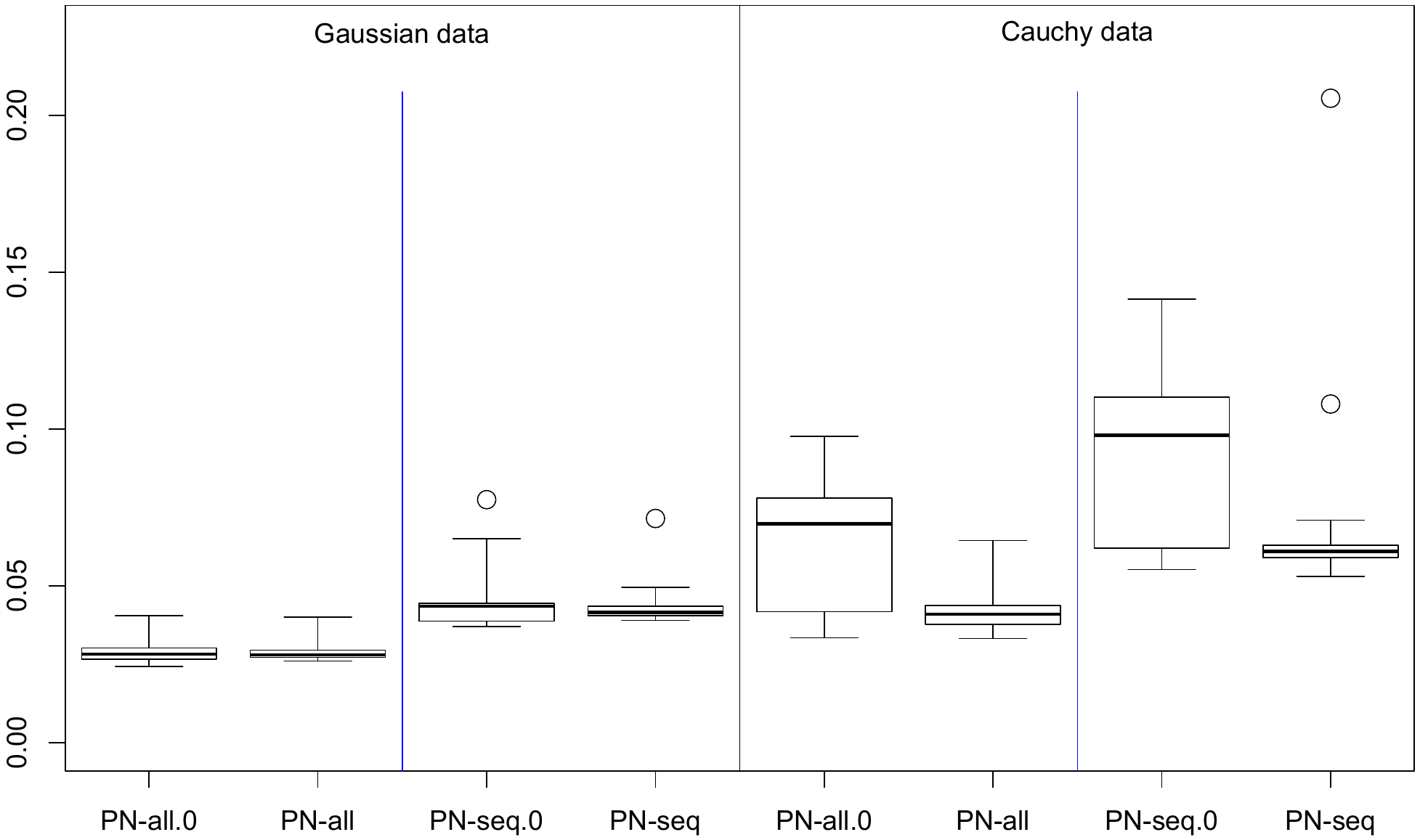}
\caption{Computation times [s] of four variants of AlgorithmPN.symm ($q = 10$, $n = 100$, $\nu = 1$).}
\label{fig:PN.symm.100}
\end{figure}

\begin{figure}
\centering
\includegraphics[width=0.9\textwidth]{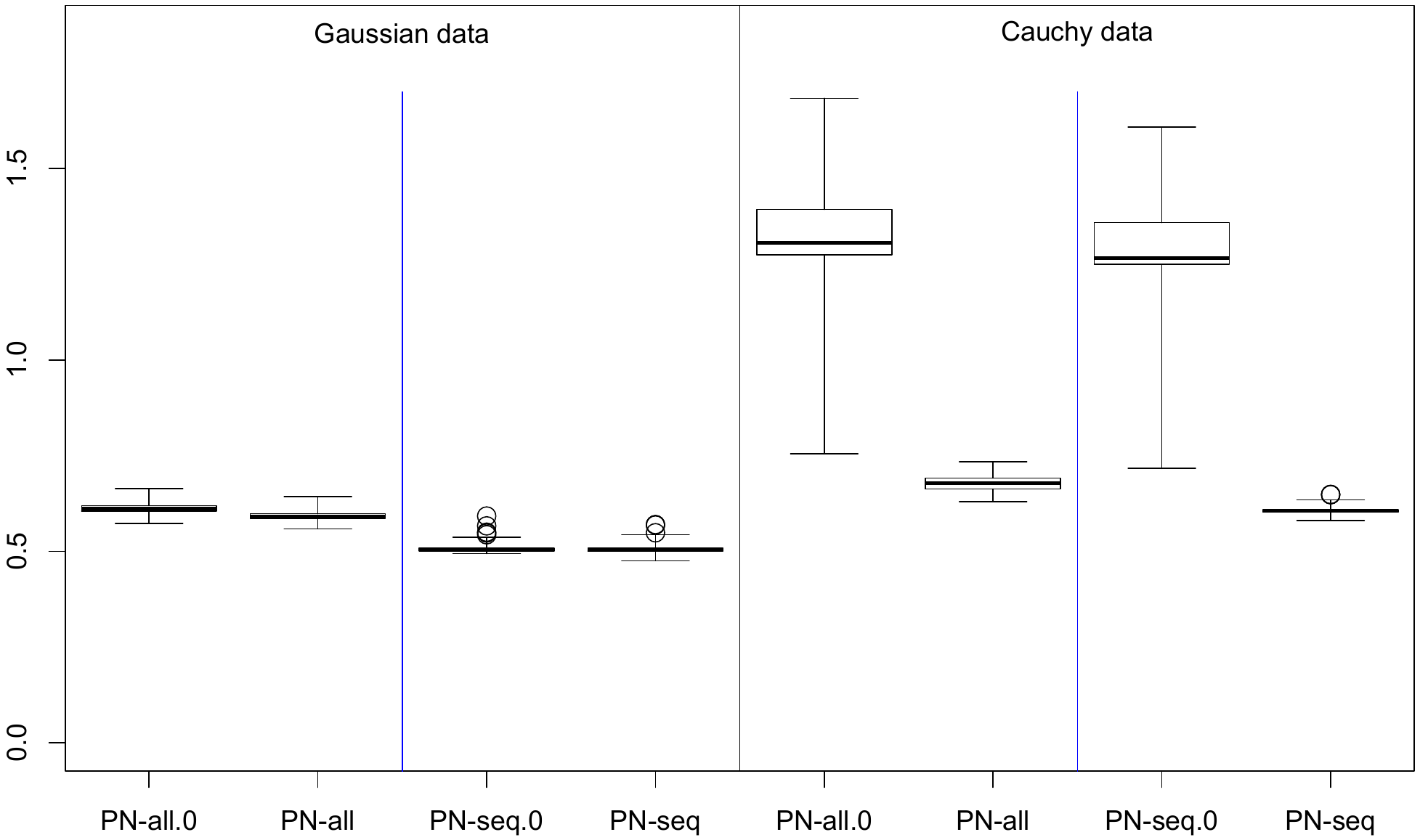}
\caption{Computation times [s] of four variants of AlgorithmPN.symm ($q = 10$, $n = 500$, $\nu = 1$).}
\label{fig:PN.symm.500}
\end{figure}

Our first simulation experiment concerns $2 \times 2$ different variants of Algorithm~PN for symmetrized estimators with $\rho = \rho_{q,1}$: On the one hand we compared storing all $N = n(n-1)/2$ pairwise differences of data vectors in a big matrix and running the algorithm in Table~\ref{tab:PN} (``PN-all'') with a less memory-intensive version where all statistics are computed sequentially as in Table~\ref{tab:PN_symm} (``PN-seq''). In both cases we first prewhitened the data by means of a scatter estimator based on $n$ randomly chosen pairs of observations, see the first four lines of pseudo-code in Table~\ref{tab:PN_symm}. On the other hand we investigated the benefits of the latter prewhitening step and implemented versions without it (``PN-all.0'' and ``PN-seq.0''). Figures~\ref{fig:PN.symm.100} and \ref{fig:PN.symm.500} show box plots of the computation times (using pure R code) for dimension $q = 10$ and sample sizes $n = 100$ and $n=500$, respectively. One sees clearly that for small to moderate sample sizes version ``PN-all'' is faster than ``PN-seq''. But for larger sample sizes ``PN-seq'' becomes clearly preferable. Comparing ``PN-all.0'' with ``PN-all'' and ``PN-seq.0'' with ``PN-seq'' shows that prewhitening is particularly beneficial for the heavy-tailed distribution and larger sample sizes. Note that all computation times for the symmetrized scatter estimators are in seconds [s] rather than milliseconds [ms] as before.

\paragraph{More efficient code.}
The new algorithms described in this paper are implemented in the R package \emph{fastM} (D\"umbgen et al.~\cite{fastM001}) which is publicly available on CRAN. This includes implementations with C++ code which are even more efficient. We did substantial simulation experiments to compare our package with other implementations of $M$-estimators, namely (i) the function \emph{cov.trob} in the package \emph{MASS} (Venables and Ripley~\cite{MASS}) and (ii) the function \emph{tM} in the package \emph{ICS} (Nordhausen et al.~\cite{ICS}). Both functions are essentially fix-point approaches. In particular, \emph{tM} is based on a maximum-likelihood and EM interpretation of the fixed point equation and uses algorithm FP$_3$ by Arslan et al.~\cite{Arslan_etal_1995} mentioned before. All in all our new algorithms were always comparable, often faster and in some settings even substantially faster than the other methods. A fair comparison is difficult, though, because the established algorithms use different stopping criteria. Both \emph{cov.trob} and \emph{tM} update the location and scatter parameters separately and do not treat it as our algorithms do, as a scatter-only problem. For the symmetrized estimator with $\rho = \rho_{0,q}$, there is the function duembgen.shape available in the R package \emph{ICSNP} (Nordhausen et al.~\cite{ICSNP}), which is essentially Algorithm~FP and utilizes R and C code.

\paragraph{Comparisons for symmetrized scatter estimators, II.}
Finally, Tables~\ref{tab:ComparisonsDuembgenN500} and \ref{tab:ComparisonsDuembgenN2000} compare the performance of the symmetrized estimator as implemented in \emph{fastM} with pure R code and with C++ code, where $\rho = \rho_{\nu,q}$, $\nu = 0,1$. The results show that Algorithm~PN with C++ code is substantially faster than its pure R version.

\begin{table}
\[
	\begin{array}{|l|c|c|c|c||c|c|c|c|}
	\cline{2-9}
	\multicolumn{1}{l}{}
	& \multicolumn{4}{|l||}{\text{Gaussian data}}
	& \multicolumn{4}{|l|}{\text{Cauchy data}} \\
	\cline{2-9}
	\multicolumn{1}{l|}{}
	& \text{Iter.} & \text{Time [s]} & \text{Time [s]} & \!\text{Rel.\ eff.}\!
	& \text{Iter.} & \text{Time [s]} & \text{Time [s]} & \!\text{Rel.\ eff.}\! \\
	\multicolumn{1}{l|}{}
	&              & \text{R}        & \text{C++}      &
	&              & \text{R}        & \text{C++}      & \\
	\cline{2-9}
	\multicolumn{9}{c}{} \\[-1.5ex]

	\cline{1-1}
	\multicolumn{1}{|l|}{\bs{\nu = 0}} \\
	\hline
	q=5
	& 4.0 \ (0) & 1.2 \ (0.3) & 0.2 \ (0.1) & \bs{6.81}
	& 5.1 \ (0) & 1.3 \ (0.2) & 0.2 \ (0.1) & \bs{5.67} \\
	\hline
	q=10
	& 5.0 \ (0) & 1.7 \ (0.4) & 0.4 \ (0.2) & \bs{3.91}
	& 6.0 \ (0) & 2.1 \ (0.4) & 0.5 \ (0.3) & \bs{4.04} \\
	\hline
	q=20
	& 5.0 \ (0) & 2.9 \ (0.7) & 0.9 \ (0.3) & \bs{3.13}
	& 6.9 \ (0) & 3.7 \ (1.0) & 1.2 \ (0.3) & \bs{3.15} \\
	\hline

	\multicolumn{9}{c}{} \\[-1.5ex]

	\cline{1-1}
	\multicolumn{1}{|l|}{\bs{\nu = 1}} \\
	\hline
	q=5
	& 4.0 \ (0) & 1.2 \ (0.3) & 0.2 \ (0.1) & \bs{6.40}
	& 5.1 \ (0) & 1.3 \ (0.2) & 0.2 \ (0.2) & \bs{5.44} \\
	\hline
	q=10
	& 5.0 \ (0) & 1.7 \ (0.4) & 0.4 \ (0.2) & \bs{3.96}
	& 6.0 \ (0) & 2.0 \ (0.4) & 0.5 \ (0.3) & \bs{3.97} \\
	\hline
	q=20
	& 5.0 \ (0) & 2.9 \ (0.8) & 0.9 \ (0.3) & \bs{3.11}
	& 6.9 \ (0) & 3.7 \ (1.0) & 1.2 \ (0.4) & \bs{3.11} \\
	\hline
	\end{array}
\]

\caption{Computation costs and relative efficiencies for symmetrized scatter ($n = 500$).}
\label{tab:ComparisonsDuembgenN500}
\end{table}

\begin{table}
\[
	\begin{array}{|l|c|c|c|c||c|c|c|c|}
	\cline{2-9}
	\multicolumn{1}{l}{}
	& \multicolumn{4}{|l||}{\text{Gaussian data}}
	& \multicolumn{4}{|l|}{\text{Cauchy data}} \\
	\cline{2-9}
	\multicolumn{1}{l|}{}
	& \text{Iter.} & \text{Time [s]} & \text{Time [s]} & \!\text{Rel.\ eff.}\!
	& \text{Iter.} & \text{Time [s]} & \text{Time [s]} & \!\text{Rel.\ eff.}\! \\
	\multicolumn{1}{l|}{}
	&              & \text{R}        & \text{C++}      &
	&              & \text{R}        & \text{C++}      & \\
	\cline{2-9}
	\multicolumn{9}{c}{} \\[-1.5ex]

	\cline{1-1}
	\multicolumn{1}{|l|}{\bs{\nu = 0}} \\
	\hline
	q=5
	& 3.2 \ (0) &  7.9 \ (1.6) &  1.9 \ (0.5) & \bs{4.03}
	& 4.0 \ (0) &  9.5 \ (1.4) &  2.3 \ (0.5) & \bs{4.06} \\
	\hline
	q=10
	& 4.0 \ (0) & 14.3 \ (2.6) &  4.3 \ (0.5) & \bs{3.30}
	& 4.6 \ (1) & 16.0 \ (3.5) &  4.9 \ (1.0) & \bs{3.27} \\
	\hline
	q=20
	& 4.0 \ (0) & 33.1 \ (7.9) & 10.1 \ (0.2) & \bs{3.28}
	& 5.0 \ (0) & 40.7 \ (8.3) & 12.2 \ (0.2) & \bs{3.33} \\
	\hline

	\multicolumn{9}{c}{} \\[-1.5ex]

	\cline{1-1}
	\multicolumn{1}{|l|}{\bs{\nu = 1}} \\
	\hline
	q=5
	& 3.2 \ (0) &  7.7 \ (1.4) &  1.9 \ (0.4) & \bs{3.99}
	& 4.0 \ (0) &  9.5 \ (1.4) &  2.4 \ (0.5) & \bs{4.00} \\
	\hline
	q=10
	& 4.0 \ (0) & 14.3 \ (2.8) &  4.4 \ (0.6) & \bs{3.24}
	& 4.7 \ (1) & 16.2 \ (3.4) &  5.0 \ (0.5) & \bs{3.25} \\
	\hline
	q=20
	& 4.0 \ (0) & 33.1 \ (7.7) & 10.1 \ (0.2) & \bs{3.27}
	& 5.0 \ (0) & 40.8 \ (7.9) & 12.3 \ (0.2) & \bs{3.32} \\
	\hline
	\end{array}
\]

\caption{Computation costs and relative efficiencies for symmetrized scatter ($n = 2000$).}
\label{tab:ComparisonsDuembgenN2000}
\end{table}

\section{Proofs}
\label{sec:proofs}

\begin{proof}[\bf Proof of Corollaries~\ref{cor:convexity} and \ref{cor:expansion}]
For $t \in \R$ define $F(t) := L(B \exp(tA) B^\top, Q)$ and $B(t) := B \exp((t/2)A)$. Note that $B(t)$ is nonsingular with $B(0) = B$. For $u \in \R$,
\begin{align*}
	F(t+u) - F(t) \
	&= \ L(B(t) \exp(uA) B(t)^\top, Q) - L(B(t)B(t)^\top, Q) \\
	&= \ L(\exp(uA),Q_{B(t)}) \\
	&= \ u G(A,Q_{B(t)}) + \frac{u^2}{2} H(A,Q_{B(t)}) + o(u^2)
\end{align*}
as $u \to 0$. Since both $G(A,Q_{B(t)})$ and $H(A,Q_{B(t)})$ are continuous in $t \in \R$, this expansion shows that $F$ is twice continuously differentiable with $F'(t) = G(A,Q_{B(t)})$ and
\[
	F''(t) \ = \ H(A,Q_{B(t)}) \
	\begin{cases}
		\ge \ 0 , \\
		> \ 0 & \text{in Setting~0 if} \ A \ne 0, \tr(A) = 0 , \\
		> \ 0 & \text{in Setting~1 if} \ A \ne 0 .
	\end{cases}
\]
In particular, $F$ is convex. It is even strictly convex unless
\[
	\begin{cases}
		A = s I_q \ \text{for some} \ s \in \R
			&\text{in Setting~0} , \\
		A = 0
			&\text{in Setting~1} .
	\end{cases}
\]

To verify Corollary~\ref{cor:expansion}, we utilize the same auxiliary function $F = F(\cdot \,|\, B,A)$ and write $L(B\exp(A)B^\top, Q) - L(BB^\top, Q)$ as
\[
	F(1) - F(0)
	\ = \ F'(0) + \int_0^1 (1 - t) F''(t) \, dt
	\ = \ G(A,Q_B) + \int_0^1 (1 - t) H(A, Q_{B(t)}) \, dt .
\]
Now let $B = \Sigma^{1/2} V$ with an orthogonal matrix $V \in \Rqq$, and define
\[
	C(t) \ := \ B(t) V^\top \ = \ \Sigma^{1/2} V \exp((t/2)A) V^\top .
\]
Then
\begin{align*}
	r(B,A) \
	&= \ \|A\|^{-2} \int_0^1 (1 - t)
		\bigl( H(A,Q_{B(t)}) - H(A,Q_B) \bigr) \, dt \\
	&= \ \|A\|^{-2} \int_0^1 (1 - t)
		\bigl( H(V^\top A V,Q_{C(t)}) - H(V^\top A V,Q_{\Sigma^{1/2}}) \bigr) \, dt , \\
	r_*(B,A) \
	&= \ \|A\|^{-2} \int_0^1 (1 - t)
		\bigl( H(A,Q_{B(t)}) - H(V^\top A V,Q_*) \bigr) \, dt \\
	&= \ \|A\|^{-2} \int_0^1 (1 - t)
		\bigl( H(V^\top A V,Q_{C(t)}) - H(V^\top A V,Q_*) \bigr) \, dt ,
\end{align*}
so $|r(B,A)| + |r_*(B,A)|$ is no larger than $3/2$ times the supremum of
\[
	\bigl| H(A',Q_{\Sigma^{1/2} V_o^{} \exp(A_o) V_o^\top}) - H(A',Q_*) \bigr|
\]
over all $A',A_o \in \Rqqsym$ with $\|A'\| \le 1$, $\|A_o\| \le \|A\|/2$ and all orthogonal matrices $V_o \in \Rqq$. But this converges to zero as $\Sigma = BB^\top \to \bs{\Sigma}(Q)$ and $A \to 0$, because then
\begin{align*}
	\bigl\| \Sigma^{1/2} V_o^{} \exp(A_o) V_o^\top - \bs{\Sigma}(Q)^{1/2} \bigr\| \
	&\le \ \|\Sigma^{1/2}\| \|V_o^{}\exp(A_o)V_o^\top - I_q\|
		+ \|\Sigma^{1/2} - \bs{\Sigma}(Q)^{1/2} \| \\
	&= \ \|\Sigma^{1/2}\| \|\exp(A_o) - I_q\|
		+ \|\Sigma^{1/2} - \bs{\Sigma}(Q)^{1/2} \| \\
	&\to \ 0 .
\end{align*}
Finally, because $G(Q_*) = I_q - \Psi(Q_*) = 0$, we may write
\begin{align*}
	H(V^\top A V, Q_*) \
	&= \ \bigl\langle (V^\top A V)^2, I_q \bigr\rangle
		+ \int \rho''(\|x\|^2) (x^\top V^\top A V x)^2 \, Q_*(dx) \\
	&= \ \|A\|^2
		+ \int \rho''(\|x\|^2) (x^\top V^\top A V x)^2 \, Q_*(dx) .
\end{align*}\\[-7ex]
\end{proof}

\begin{proof}[\bf Proof of Theorem~\ref{thm:convergence}]
Dropping the index $k$ for the moment, suppose that $\Sigma = BB^\top$ is our current candidate parameter. Then one step of Algorithm~FP replaces $\Sigma$ with
\[
	B \Psi(Q_B) B^\top \ = \ \int \rho'(x^\top \Sigma^{-1}x) xx^\top \, Q(dx) .
\]
Hence $L(\Sigma,Q)$ changes by
\[
	\delta_1(\Sigma) \ := \ L(B \Psi(Q_B) B^\top, Q) - L(\Sigma,Q)
	\ = \ L(\Psi(Q_B), Q_B) \ \le \ 0 ,
\]
and the inequality is strict unless $\Sigma$ minimizes $L(\cdot,Q)$ already, see \eqref{eq:PsiQ.better.than.I}. Note also that $\delta_1(\Sigma)$ is a continuous function of $\Sigma$.

Algorithm~PN is slightly more difficult to quantify, because the eigenmatrix $U$ in the representation $\Psi(Q_B) = U \diag(\phi) U^\top$ is not unique. However,
\begin{align*}
	\min_{a \in \R^q} \Bigl( \tilde{G}(Q_{BU})^\top a
		+ \frac{1}{2} a^\top \tilde{H}(Q_{BU}) a \Bigr) \
	&\le \ \min_{a \in \mathrm{span}(\tilde{G}(Q_{BU}))} \Bigl( \tilde{G}(Q_{BU})^\top a
		+ \frac{1}{2} a^\top \tilde{H}(Q_{BU}) a \Bigr) \\
	&= \ \frac{- \|\tilde{G}(Q_{BU})\|^2}
		{2 \tilde{G}(Q_{BU})^\top \tilde{H}(Q_{BU}) \tilde{G}(Q_{BU})} \\
	&= \ \frac{- \|G(Q_{BU})\|^2}
		{2 H(G(Q_{BU}), Q_{BU})} \\
	&= \ \frac{- \|G(Q_{\Sigma^{1/2}})\|^2}
		{2 H(G(Q_{\Sigma^{1/2}}), Q_{\Sigma^{1/2}})} .
\end{align*}
In the last step we utilized that fact that $BU = \Sigma^{1/2} W$ for some orthogonal matrix $W \in \Rqq$, and that $G(Q_{BU}) = W^\top G(Q_{\Sigma^{1/2}}) W$, $H(G(Q_{BU}),Q_{BU}) = H(G(Q_{\Sigma^{1/2}}),Q_{\Sigma^{1/2}})$. Consequently, the change of $L(\Sigma,Q)$ with Algorithm~PN is at least
\[
	\delta_2(\Sigma) \ := \ \max \Bigl( \delta_1(\Sigma),
		\frac{- \|G(Q_{\Sigma^{1/2}})\|^2}
			{4 H(G(Q_{\Sigma^{1/2}}), Q_{\Sigma^{1/2}})} \Bigr) \ \le \ 0 ,
\]
again a continuous function of $\Sigma$, and the inequality is strict unless $\Sigma$ minimizes $L(\cdot,Q)$.

In Setting~1, the minimizer $\bs{\Sigma}_\rho(Q)$ is unique, and we may utilize the following standard arguments: Suppose that $(\Sigma_k)_k$ does not converge to $\bs{\Sigma}_\rho(Q)$. We know that $L(\Sigma_k,Q)$ is decreasing in $k \ge 0$, and all $\Sigma_k$ belong to the compact set $\{\Sigma: L(\Sigma,Q) \le L(\Sigma_0,Q)\}$. Hence there would exist a subsequence $(\Sigma_{k(\ell)})_\ell$ with limit $\Sigma_* \ne \bs{\Sigma}_\rho(Q)$. But then continuity of $L(\cdot,Q)$ and $\delta_j(\cdot)$ would imply that
\begin{align*}
	L(\Sigma_*,Q) \
	&= \ \lim_{\ell \to \infty} L(\Sigma_{k(\ell)},Q) \\
	&= \ \lim_{\ell \to \infty} L(\Sigma_{k(\ell)+1},Q) \\
	&\le \ \lim_{\ell \to \infty} \bigl( L(\Sigma_{k(\ell)},Q)
		+ \delta_j(\Sigma_{k(\ell)}) \bigr) \\
	&= \ L(\Sigma_*,Q) + \delta_j(\Sigma_*) \\
	&< \ L(\Sigma_*,Q) .
\end{align*}

In Setting~0, note first that $L(\Sigma,Q)$, $\Psi(Q_B)$ and $H(Q_B)$ remain unchanged if we replace $(\Sigma, B)$ with $(t\Sigma,t^{1/2} B)$ for some number $t > 0$. Hence, with the same arguments as in Setting~1, we may conclude that $t_k \Sigma_k \to \bs{\Sigma}_0(Q)$ as $k \to \infty$, where $t_k := \det(\Sigma_k)^{-q/2}$.

Now in case of Algorithm~FP an elementary calculation shows that the matrices $M_k := \bs{\Sigma}_0(Q)^{-1/2} \Sigma_k \bs{\Sigma}_0(Q)^{-1/2}$ satisfy the equation
\[
	M_{k+1} \ = \ \int \frac{q}{x^\top M_k^{-1} x} \, xx^\top \,
		Q_{\bs{\Sigma}_0(Q)^{1/2}}^{}(dx) .
\]
Together with the equation $\Psi(Q_{\bs{\Sigma}_0(Q)^{1/2}}) = I_q$ this implies that
\[
	 \lambda_{\rm min}(M_{k+1}) \ \ge \ \lambda_{\rm min}(M_k)
	 \quad\text{and}\quad
	 \lambda_{\rm max}(M_{k+1}) \ \le \ \lambda_{\rm max}(M_k) .
\]
Hence the sequence $(M_k)_k$ converges to a multiple of the identity matrix. In other words, $(\Sigma_k)_k$ converges to a multiple of $\bs{\Sigma}_0(Q)$.

The definition of Algorithm~PN implies that for sufficiently large $k$, the new candidate $\Sigma_{k+1}$ is given by $B_k^{} \exp(\diag(a_k^{})) B_k^\top$ with $a_k \in \R^q$ satisfying $1_q^\top a_k = 0$. Hence $\det(\Sigma_{k+1}) = \det(\Sigma_k)$ for sufficiently large $k$. Consequently $(\Sigma_k)_k$ converges to a multiple of $\bs{\Sigma}_0(Q)$.
\end{proof}

\begin{proof}[\bf Proof of Lemma~\ref{lem:PN.is.right}]
The fact that $\bs{\Sigma}(Q)$ is a positive multiple of $\Sigma_o$ follows from simple equivariance considerations as outlined in \cite{Duembgen_etal_2015}. Now let $\bs{\Sigma}(Q) = CC^\top$ with nonsingular $C \in \Rqq$, and let $Z := C^{-1} X$ with $X \sim Q$. The random vector $Z$ has a spherically symmetric distribution around $0$ in the sense that for any orthogonal matrix $V \in \Rqq$, the distributions of $V^\top Z$ and $Z$ coincide. We may write
\begin{align*}
	\Psi(Q_B) \
	&= \ \Ex \bigl[ \rho'(\|B^{-1}X\|^2) (B^{-1}X)(B^{-1}X)^\top \bigr] \\
	&= \ B^{-1}C
		\Ex \bigl[ \rho'(Z^\top C^\top \Sigma^{-1} C Z) ZZ^\top \bigr]
		C^\top B^{-\top} .
\end{align*}
Next let
\[
	C^\top \Sigma^{-1} C \ = \ V \diag(\gamma) V^\top
\]
with an orthogonal matrix $V \in \Rqq$ and a vector $\gamma \in (0,\infty)^q$ containing the eigenvalues of $C^\top \Sigma^{-1} C$, i.e.\ the eigenvalues of $\Sigma^{-1} \bs{\Sigma}(Q)$. Then
\[
	B^{-1} C \ = \ \tilde{U} \diag(\gamma)^{1/2} V^\top
\]
for another orthogonal matrix $\tilde{U}$, so
\begin{align*}
	\Psi(Q_B) \
	&= \ \tilde{U} \diag(\gamma)^{1/2} V^\top
		\Ex \bigl[ \rho'(Z^\top V \diag(\gamma) V^\top Z) ZZ^\top \bigr]
			V \diag(\gamma)^{1/2} \tilde{U}^\top \\
	&= \ \tilde{U} \diag(\gamma)^{1/2}
		\Ex \bigl[ \rho'((V^\top Z)^\top \diag(\gamma) (V^\top Z)) (V^\top Z) (V^\top Z)^\top \bigr]
			\diag(\gamma)^{1/2} \tilde{U}^\top \\
	&= \ \tilde{U} \diag(\gamma)^{1/2}
		\Ex \bigl[ \rho'(Z^\top \diag(\gamma) Z) ZZ^\top \bigr]
			\diag(\gamma)^{1/2} \tilde{U}^\top \\
	&= \ \tilde{U} \diag(\gamma)^{1/2}
		\Ex \Bigl[ \rho' \Bigl( \sum_{i=1}^q \gamma_i Z_i^2 \Bigr) (Z_jZ_k)_{j,k=1}^q \Bigr]
			\diag(\gamma)^{1/2} \tilde{U}^\top \\
	&= \ \tilde{U} \diag(\gamma)^{1/2}
		\Ex \Bigl[ \rho' \Bigl( \sum_{i=1}^q \gamma_i Z_i^2 \Bigr)
			\diag \bigl( (Z_j^2)_{j=1}^q \bigr) \Bigr]
			\diag(\gamma)^{1/2} \tilde{U}^\top \\
	&= \ \tilde{U}
		\Ex \Bigl[ \rho' \Bigl( \sum_{i=1}^q \gamma_i Z_i^2 \Bigr)
			\diag \bigl( (\gamma_j Z_j^2)_{j=1}^q \bigr) \Bigr]
			\tilde{U}^\top ,
\end{align*}
by spherical symmetry of the distribution of $Z$. Hence
\[
	\Psi(Q_B) \ = \ \tilde{U} \diag(\phi) \tilde{U}^\top
\]
with $\phi \in (0,\infty)^q$ given by
\[
	\phi_j
	\ := \ \Ex \Bigl( \rho' \Bigl( \sum_{i=1}^q \gamma_i Z_i^2 \Bigr) \gamma_j Z_j^2 \Bigr) .
\]
Moreover, since $\rho' > 0$ and the distribution of $(Z_i^2)_{i=1}^q$ is invariant under permuting the components of $Z$,
\[
	\phi_j = \phi_k \ \ \text{if, and only if,} \ \ \gamma_j = \gamma_k .
\]

One may also say that $\phi$ is the unique vector of eigenvalues of $\Psi(Q_B)$, and the columns $\tilde{u}_1, \tilde{u}_2, \ldots, \tilde{u}_q$ of $\tilde{U}$ are corresponding eigenvectors. If we consider another spectral decomposition $\Psi(Q_B) = U \diag(\phi) U^\top$ with $U$ having orthonormal columns $u_1, u_2, \ldots, u_q$, then
\[
	U \exp(\diag(a)) U^\top \ = \ \tilde{U} \exp(\diag(a)) \tilde{U}^\top
\]
for any vector $a \in \R^q$ such that $a_j = a_k$ whenever $\phi_j = \phi_k$. In particular, if we choose $a := \bigl( \log(\gamma_j) \bigr)_{j=1}^q$, then
\begin{align*}
	B U & \exp(\diag(a)) U^\top B^\top \\
	&= \ B \tilde{U} \diag(\gamma) \tilde{U}^\top B^\top
		\ = \ B (B^{-1} C) (B^{-1} C)^\top B^\top
		\ = \ CC^\top
		\ = \ \bs{\Sigma}(Q) .
\end{align*}\\[-7ex]
\end{proof}

\paragraph{Acknowledgement.}
The authors are grateful to Mathias Drton for his interest and questions which led to Lemma~\ref{lem:PN.is.right}. We are also indebted to an anonymous referee for detailed and constructive comments.



\end{document}